\documentclass[%
 reprint,
 amsmath,amssymb,
 aps,
prb,
]{revtex4-2}

\usepackage{graphicx}
\usepackage{dcolumn}
\usepackage{bm}
\usepackage{placeins}
\usepackage{float}
\usepackage{svg}
\usepackage{afterpage}
\usepackage{multirow} 

\begin{document}

\preprint{APS/123-QED}

\title{Analysis of Josephson Junction Barrier Variation: a Combined  Electron Microscopy, Breakdown and Monte-Carlo Approach}

\author{Oscar~W.~Kennedy}\email{okennedy@oqc.tech}
\author{Kevin~G.~Crawford}
\author{Kowsar~Shahbazi}
\author{Connor~D.~Shelly}\email{cshelly@oqc.tech}

\affiliation{%
 Oxford Quantum Circuits, Thames Valley Science Park, Shinfield, Reading, United Kingdom, RG2 9LH}%


\begin{abstract}
Josephson junctions manufactured to tight tolerances are necessary components for superconducting quantum computing. Developing precise manufacturing techniques for Josephson junctions requires an understanding of their make-up and robust feedback metrics against which to optimise. 
Here we consider complementary techniques assessing what conclusions they allow us to draw about the barriers in junctions. 
Monte-Carlo simulations of barriers show that standard deviations of 15-20\% of the total barrier thickness are compatible with our experimental data.
Electrical breakdown allows us to probe the weakest points in barriers. Narrowing the distribution of this breakdown provides a promising feedback mechanism for barrier optimisation. 
Grouping junctions by breakdown voltage allows us to identify sub-ensembles of junctions with different median resistance. 
Transmission electron microscopy can be used to find average barrier thickness, although we highlight challenges forming robust conclusions on the distribution of thicknesses in a barrier from these experiments. 


\end{abstract}

\maketitle


\section{Introduction}
Josephson junctions (JJs) are ubiquitous to superconducting quantum computers~\cite{arute2019quantum} and other superconducting circuits such as parametric amplifiers~\cite{macklin2015near} and therefore have significant technological importance. 
The most widely used type of JJ in superconducting technologies is based on an insulating layer of AlO$_{\rm x}$ which separates two superconducting electrodes. The AlO$_{\rm x}$ layer is formed by oxidising the surface of a layer of metallic aluminum before covering the AlO$_{\rm x}$ layer with a second superconducting electrode, also aluminum. This creates a barrier which is (i) amorphous (ii) non-stoichiometric, and (iii) with thickness that varies across the junction~\cite{nik2016correlation, fritz2019structural, zeng2016atomic, lapham2022computational}. 
Developing manufacturing techniques for JJs to be made with fine tolerances requires a detailed understanding of the junction, as well as robust feedback metrics against which manufacturing techniques can be optimised. 

Typically, when designing superconducting devices reliant on JJs, a specific critical current value, $I_{\rm C}$, is targeted. Ensemble measurements of $I_{\rm C}$ are typically used to characterize Nb/AlO$_{\rm x}$/Nb JJ properties~\cite{yohannes2005characterization, west2022wafer, semenov2019very}. This property can be harder to directly measure across statistically significant ensembles in Al/AlO$_{\rm x}$/Al JJs due to their critical temperature being substantially below liquid helium temperatures. 
A proxy for this value in junctions with clean insulating barriers is the room temperature resistance~\cite{kreikebaum2020improving, ambegaokar1963tunneling} and resistance tolerances are typically used to characterise manufacturing processes. These processes are in turn optimised considering feedback from resistance measurements and techniques such as scanning electron microscopy~\cite{osman2021simplified, moskalev2023optimization}. 
Consequently many problems with repeatable JJ manufacture have been solved and control of the area of JJs is now much improved~\cite{kreikebaum2020improving, pishchimova2023improving}. As a result the current state of the art in as-fabricated JJ resistance spread is $\sim$2-3\%~\cite{osman2023mitigation, acharya2024integration, van2024advanced}. Once lateral dimensions of junctions are perfectly controlled the remaining variation between junctions must be attributed to differences between their barriers although it is not clear what level of resistance variation this causes in typical manufacturing processes.

Variability in junction barriers has more complex consequences on the junction physics than being a poorly controlled variable which rescales the critical current affecting manufacturing tolerance.
Non-ideal JJs can be characterised by their expanded current phase relationship (C$\phi$R). In these realistic junctions, the first Josephson equation, $I = I_C \sin(\phi)$ where $\phi$ is the superconducting phase drop across the JJ~\cite{josephson1962possible}, becomes more complicated with current given by the Fourier series $I = \sum_n{I_{C,n} \sin(n\phi)}$~\cite{golubov2004current}. Imperfections in the barrier cause variations in $I_{C, n}$, changing the critical current and adding higher order terms to the C$\phi$R~\cite{bayros2024influence}. These higher order terms can then affect the properties of devices based on these JJs~\cite{willsch2024observation}. 

Variation in dielectric thin films, like the barrier, is a well studied problem with relevance to devices including field effect transistors~\cite{lee2006gate}, metal-insulator-semiconductor or metal-insulator-metal structures~\cite{buchanan1997reliability, ekanayake2004metal} and magnetic tunnel junctions~\cite{da2000tunneling}. 
The barriers in JJs are typically $\sim1-2$~nm~\cite{zeng2015direct} so even a single O$^{2-}$ ion represents a substantial fraction of the average barrier thickness with ionic radius of $\sim$0.14~nm~\cite{dobrovinskaya2009properties}.
Generating a full 3D map of barriers at this resolution isn't yet possible although various techniques have been used to improve the understanding of barriers. 
(Scanning) Transmission Electron Microscopy ((S)TEM) and associated techniques give chemical and structural information at sub-nanometre resolution~\cite{fritz2019structural, liu2023unveiling}. 
Atom Probe Tomography results in 3D reconstructions of the barrier although to date, results on AlO$_{\rm x}$ barriers do not show the requisite resolution~\cite{supple2021atomic, liu2023unveiling}. 
Scanning Tunneling Microscopy (STM) maps the as-grown oxide prior to being capped by a final metal layer~\cite{da2000tunneling} and oxygen migration occurs after this final capping~\cite{cyster2021simulating}.
Simulations of the growth process by molecular dynamics give insight into atomic structures in typical barriers~\cite{cyster2021simulating}.
Studies of breakdown voltages of barriers are used in device communities, such as  field effect transistors, to probe device-to-device variation in barrier weak points~\cite{palumbo2020review} although have not seen substantial use in JJs.

In this work we consider electrical measurements and STEM imaging of barriers, discussing what conclusions can be robustly drawn from these types of characterisation. We extend the typical range of electrical measurements to sample non-linear parts of the current-voltage (IV) curves and junction breakdown. We complement this by Monte-Carlo simulations of IVs and breakdown resulting from non-uniform barrier thickness. We are able to recreate the extended measurements with these Monte-Carlo simulations considering only statistically distributed thicknesses within single junctions and find that skewed thickness distributions fit our experiments best. 
We then consider how thickness information is derived from STEM imaging, a technique that has previously been used to measure thickness distributions in barriers. We consider how choices in image processing and non-idealities in the barrier electrodes may affect conclusions drawn from this type of analysis. 

We study tunnel barriers made by shadow angle evaporation~\cite{dolan1977offset} of aluminum forming Al/AlO$_{\rm x}$/Al tunnel barriers which, when cooled to cryogenic temperatures, form Josephson junctions. 
Using the same processes as in ~\cite{acharya2024integration}, junctions are defined in a Dolan bridge geometry ~\cite{dolan1977offset} via electron beam lithography. A base layer of aluminum is evaporated onto a sapphire substrate and is then oxidised in a static oxidation process before being capped with a top layer of aluminum and lifted off. All junctions in this work are fabricated on a single sapphire wafer in one fabrication run. 

\section{Electrical measurements}

\begin{figure}
    \centering
    \includegraphics[width=\linewidth]{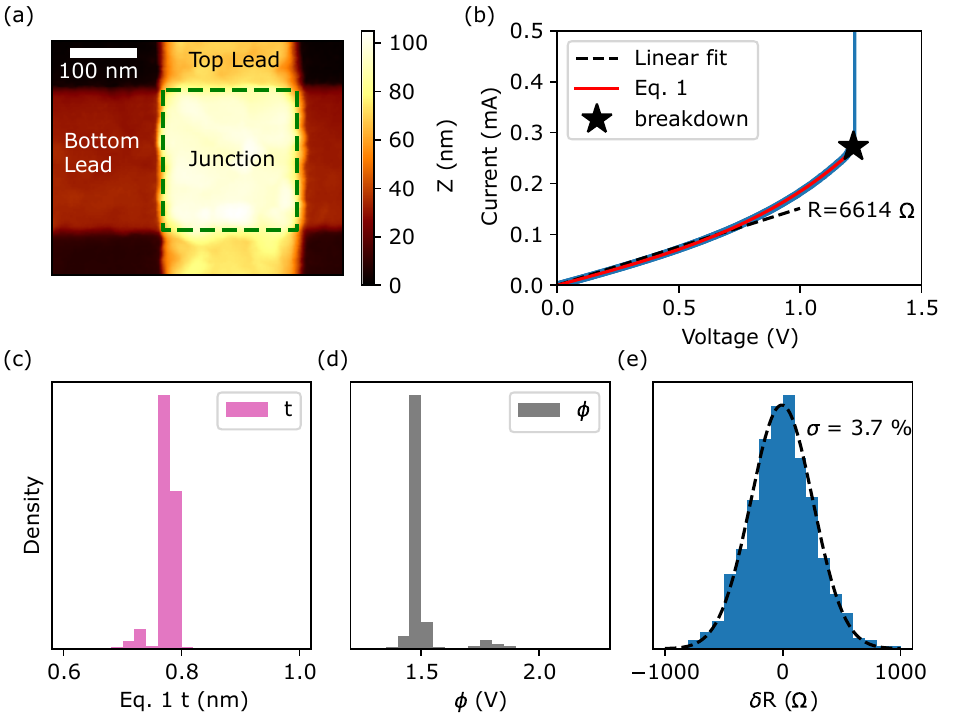}
    \caption{(a) An AFM measurement of a JJ from the ensemble of JJs (b) An IV of a typical junction showing the extraction of the breakdown voltage, resistance from a linear fit to the low voltage regime and a fit to the full Simmons model. (c,d) Results from analysing an ensemble of 598 junctions fabricated on a 3'' wafer. (c) A histogram showing the outputs from fitting Eq.~\ref{eq:iv} to measured IVs with thickness, barrier height and nominal area shown on the same x axis. (d) A histogram of the resistance deviation from the median resistance of the junction ensemble with a median resistance value of 7122~$\Omega$. }
    \label{fig:dc_meas}
\end{figure}

Fig.~\ref{fig:dc_meas} shows results from fitting a set of 597 IV curves. These are collected from junctions manufactured with the same target dimensions. An atomic force microscope (AFM) measurement of one of these junctions is shown in Fig.~\ref{fig:dc_meas}~(a), from which we find the junction dimensions to be $\sim240 \times 240$~nm = 5.76$\times 10^4$~nm$^2$.
We apply a voltage across each junction which is swept from 0~V to a value up to 1.6~V as shown in Fig.~\ref{fig:dc_meas}~(b). At low voltages the IVs of the JJs are linear and we fit this dependence at voltages below 0.02~V to extract the junction resistance. At voltages below 1.6~V, each junction fails at its breakdown voltage and becomes an ohmic channel with much lower resistance (typically hundreds of Ohms). 

We also fit the IVs for all voltages below the breakdown voltage using the Simmons model, which gives the current-voltage properties for a thin rectangular tunnel junction as a function of the potential barrier height and its thickness~\cite{simmons1963generalized};
\begin{multline}
    I = \frac{Ae^2}{2\pi h t^2} \Biggl[ \left(\phi - \frac{V}{2}\right)\exp \left(-K\sqrt{\phi - \frac{V}{2}} \right) - \\ 
    \left(\phi + \frac{V}{2}\right)\exp \left(-K\sqrt{\phi + \frac{V}{2}} \right) \Biggr]
    \label{eq:iv}
\end{multline}
where $A$ is the junction area (constrained to the value found by AFM in Fig.~\ref{fig:dc_meas}~a), $t$ the junction thickness, $\phi$ the barrier height and $K = 4\pi t\sqrt{2m_ee}/h$. Fitting this model gives a single-valued thickness which, as discussed in the introduction and shown in the literature, for example by STM~\cite{da2000tunneling}, is a simplification of the true barrier in these junctions. The simple analytical form makes it computationally inexpensive to compute, which is important when used in Monte-Carlo simulations shown later.

In Fig.~\ref{fig:dc_meas}~(c, d) we show the results of fitting Eq.~\ref{eq:iv} presenting $t$ and $\phi$ respectively. We find median $\pm$ standard deviation values of the fit parameters given by $t$~=~0.78~nm $\pm 3\%$, $\phi = $1.48~eV $\pm 6\%$. In  Fig.~\ref{fig:dc_meas}~(e) we show a histogram of the JJ resistances extracted from the linear fit, centred at 7122~$\Omega$ with a normally distributed resistance and a standard deviation of 3.7~\%. 
We use these experimental results to define targets for Monte-Carlo simulations of IV curves. 

We perform Monte-Carlo simulations where we allow thickness to vary across a junction trying to understand if we can recreate experimental measurements considering only thickness variations within junctions. 
To build these simulations we define a junction with an area corresponding to the overlap between the two electrodes in a JJ (shown in Fig.~\ref{fig:dc_meas}~a and the supplemental materials). This area is divided into equally sized pixels which act as parallel conduction channels. The thickness of the barrier at each pixel is drawn from a statistical distribution and then Eq.~\ref{eq:iv} is computed as the IV for that pixel. Combining the IVs of each pixel as parallel conduction paths, simulates the IV of the whole junction. We consider both normally distributed and log-normally distributed thicknesses, where a log-normal distribution is an example of a skewed thickness distribution.     

\begin{figure*}
    \includegraphics[width=\linewidth]{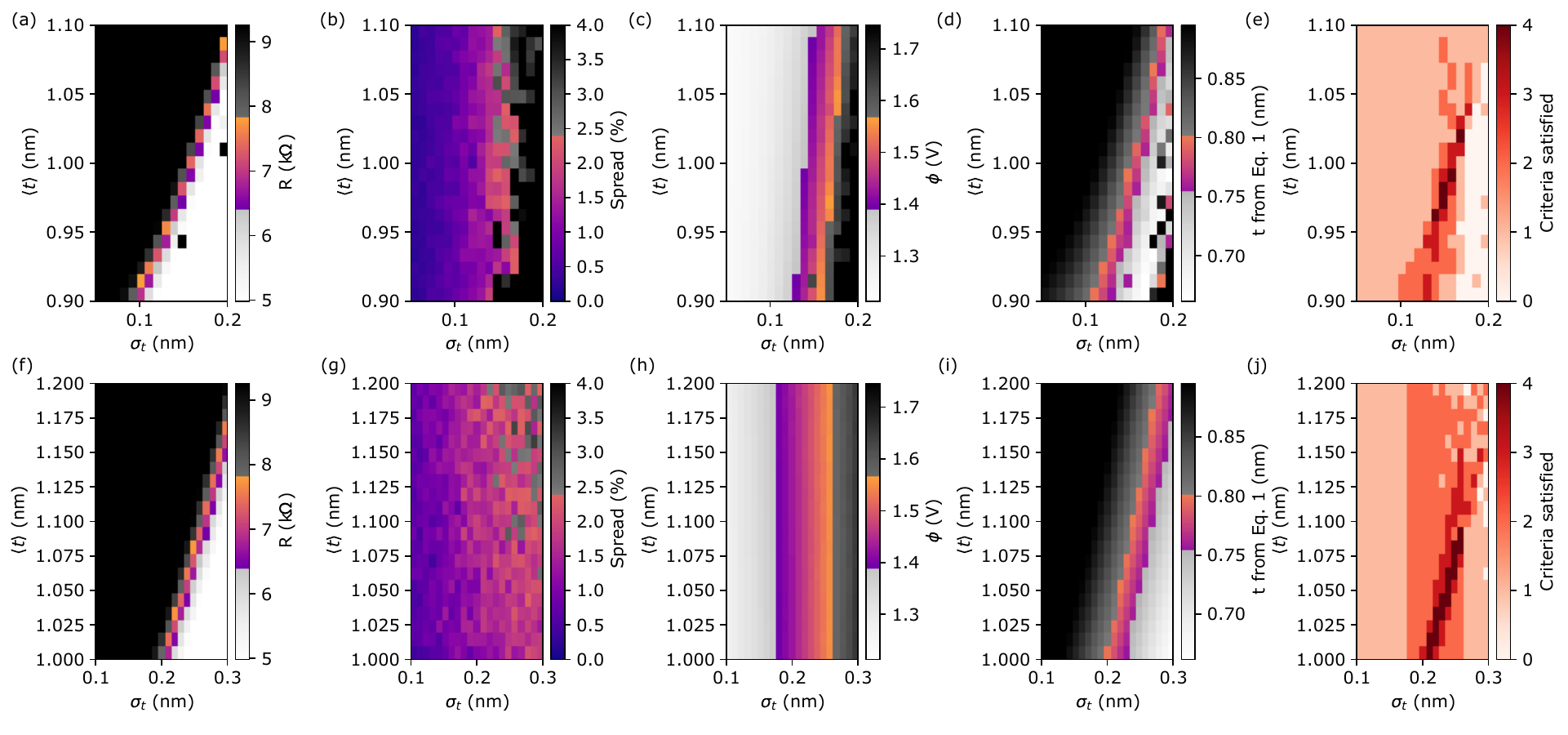}
    \caption{Results from Monte-Carlo simulations of barriers with normally (a - e) and log-normally (f - j) distributed thicknesses. Heatmaps of (a, f) resistance, (b, g) resistance spread (c, h) barrier height ($\phi$), (d, i) fitted thickness. Resistance and its spread are found by fitting a straight line to the low voltage region of the simulated IV and other parameters are from found from fitting Eq.~\ref{eq:iv} to the full voltage range of the simulated IV. Any regions within the experimental variation (or 10\% instead for resistance) are shown using a purple to orange heatmap, whereas values outside this range are shown in gray-scale (e,j) For each pixel we sum the number of experimental criteria matched by the simulation. Dark hues indicate regions where the experiment and simulation coincide, indicating that there is a parameter regime for each distribution where this matching is achieved. }
    \label{fig:mciv}
\end{figure*}

We characterise each simulated IV curve by refitting it to Eq~\ref{eq:iv} which allows us to compare simulated IVs to experimental IVs. In the fitting routine we constrain the area of the junction to be the same value as that found by AFM. Fitting the linear regime of the IV gives the resistance of the junction. 
Repeating this protocol for multiple junctions (20 in this work) gives the junction-to-junction variation for a given thickness distribution. 
The pixel size is an important parameter in these simulations with smaller pixels giving less junction-to-junction variation. Pixels less than $\sim$0.2~nm become unphysical as they drop below ionic radii of the barrier~\cite{dobrovinskaya2009properties} giving a lower bound on pixel size. 
We perform these simulations with a pixel size of 1~nm and a barrier height of $\phi = 1.22$~eV. 
The pixel size is chosen to correspond approximately to the thickness of barriers assuming that the barrier thickness doesn't change on a length-scale much less than the thickness and also as a compromise for computational speed.
The value of barrier height is important for inferring accurate thicknesses from any subsequent modeling. Literature values vary with changes in barrier height being ascribed to the termination of the oxide barrier (Al/O) and the crystal structure of the underlying metal~\cite{koberidze2016effect, koberidze2018structural}. Experimental studies also using the Simmons model identify barrier heights ranging from $\sim$0.9~eV to $\sim$1~eV~\cite{dorneles2003use}. Density functional theory (DFT) simulations of amorphous Al/AlO$_{\rm x}$/Al barriers show dependence on roughness of the underlying metal and find effective barrier heights ranging from 0.2~eV to 1~eV~\cite{kim2020density}. 

We show the results of these Monte-Carlo simulations in Fig.~\ref{fig:mciv}. We compare these to experimental results with different criteria for matching to experimental results. Resistance is considered a match if the simulated value is within 10\% of experimental values. Resistance spread is considered a match if the spread is $\leq$~2.4\%, as presented in the supplemental information of Ref.~\cite{acharya2024integration} for junctions made with this recipe. Simulated values of $t$ and $\phi$ are considered to match if they are within one standard deviation of the median value found by fitting ensembles of experimental IV curves as shown in Fig.~\ref{fig:dc_meas}. 
The top and bottom rows show results for normally distributed and log-normally distributed barrier thicknesses respectively. In each heatmap we colour any cells compatible with our experimental with a purple - orange heatmap, and show cells incompatible with experimental results in greyscale. For both distributions there is a region where the experimental results shown in Fig.~\ref{fig:dc_meas} are matched well by simulations indicated by the darkest hues in Fig.~\ref{fig:mciv}~(e,j). 
Using a multi-valued thickness within a junction we are able to match experimental IVs whilst maintaining the known lateral dimensions of the junctions.

\begin{figure}
    \centering
    \includegraphics[width=\linewidth]{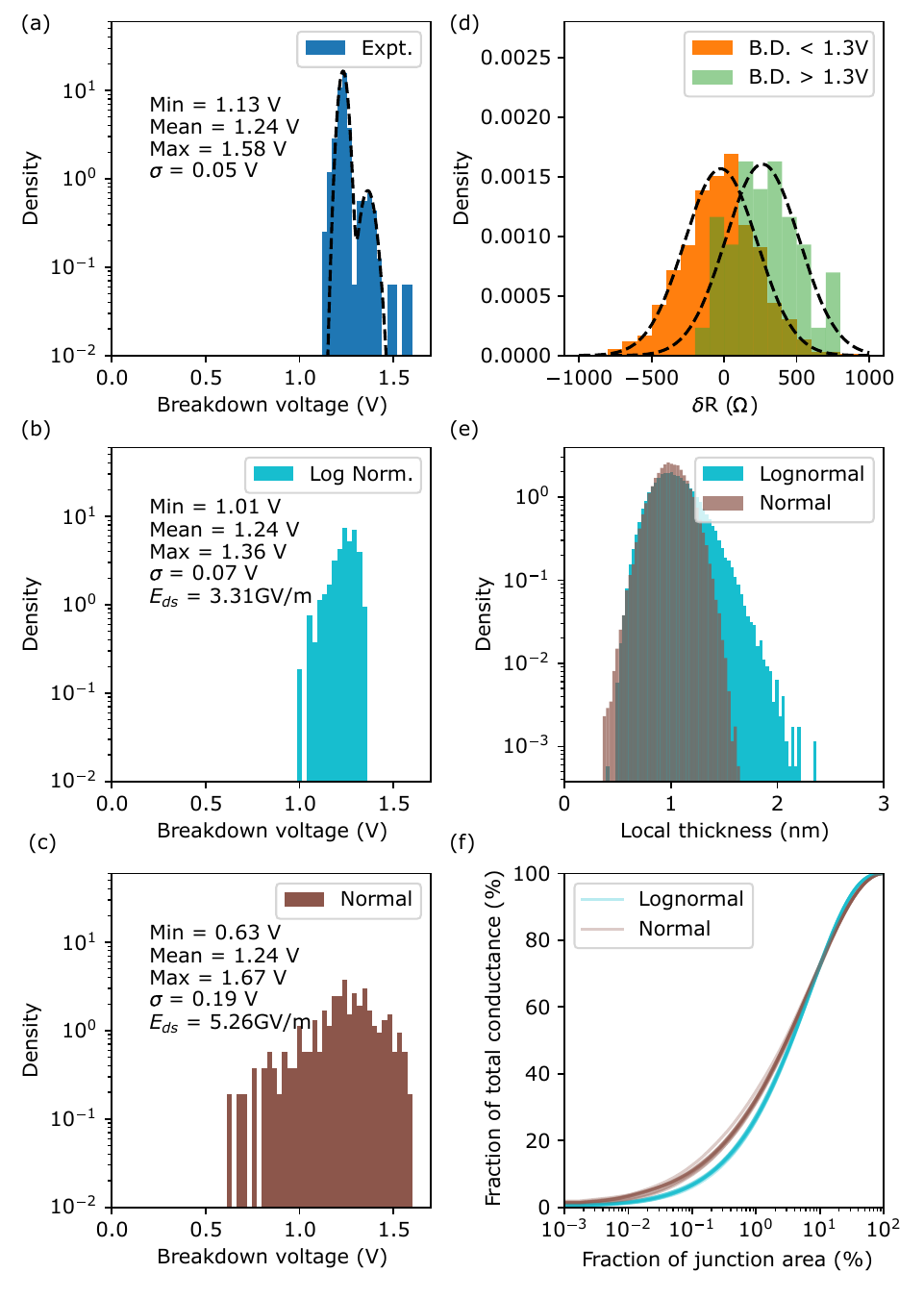}
    \caption{Comparison of (a) experimentally measured breakdown voltages with simulated breakdown voltages for (b) log-normally and (c) normally distributed barriers created with distributions in regions where all experimental criteria are matched in Fig.~\ref{fig:mciv}. (a) Shows a double-Gaussian fit indicating a bimodal breakdown distribution. 
    Simulated distributions are found by taking the thinnest points in randomly generated barriers with a mesh size of 0.2~nm and rescaling this by a dielectric strength ($E_{\rm ds}$) to keep the mean breakdown voltage equal to the experimental values. 
    (d) The resistance of junctions distinguishing junctions with breakdown above/below the midpoint of the bimodal distribution 1.3~V, shown by the black dashed line in (a). 
    (e) A histogram of the full thickness distribution of example junctions generated with the parameters from (b, c). 
    (f) Cumulative conductance of junctions from 10 different junctions generated with parameters from (e) as a fraction of the total junction area. }
    \label{fig:breakdown_stats}
\end{figure}

These simulations also show that inferring the barrier height by extracting the value from fits to the Simmons model is not straightforward. 
In Fig.~\ref{fig:mciv} we use a single value of barrier height $\phi = 1.22$~eV. However, the value of $\phi$ extracted from refitting these simulations shown in Fig.~\ref{fig:mciv}~(c,h) deviates from this value, depending on the underlying thickness distribution. 
For instance, moving along the x axis for fixed value of $\langle t \rangle$ in both Fig.~\ref{fig:mciv}~(c,h) the barriers with greater thickness variation appear to have a larger barrier heights. 
This may explain observations such as those seen in Ref.~\citenum{dorneles2003use} where they conclude that oxidation time has changed the barrier height. Instead of changing barrier height, it may be that the oxidation time is changing thickness distributions (either average or variation), which appear as potential barrier heights when interpreted via fitting to the Simmons model. 
Given the variation in literature values and imperfect inference from electrical measurements, direct measurements of the barrier height would be advantageous, for instance by x-ray photoelectron spectroscopy techniques such as in Ref.~\cite{snijders2002structure}.

To understand the effect of changing barrier height, we perform these simulations for $\phi$ = 1.5~eV, 1~eV and 0.8~eV and show these in the supplemental information and compare the results of fitting the Simmons model to data. With large barrier heights (similar to the median values found experimentally) we find a parameter regime where simulations match experiment. This parameter regime has very uniform barriers (barriers which vary less than an ionic radius). This is contrary to observations such as those in Ref.~\cite{fritz2019optimization, oh2025correlating} where thickness variation can be seen at grain boundaries. 
For lower barrier heights, we do not find any regions where all parameters match experiments. The parameters which match the most experimentally imposed conditions imply a thicker average barrier and also a larger thickness standard deviation. When using the barrier height of 0.8~eV we find average thicknesses which matches our best inference from STEM imaging discussed below. This places a lower bound on barrier heights compatible with our results.
The use of a rectangular potential barrier will give a systematic offset implying a thinner barrier relative to what would be inferred using more realistic rounded potential barriers. A detailed understanding of potential barrier shapes or atomistic computations of tunneling currents would improve this model and likely result in better quantitative agreement between simulated potential barrier thicknesses and the thicknesses of AlO$_{\rm x}$ barriers. 
We could also extend this model to also draw $\phi$ from a distribution of values indicating compositional fluctuations at the expense of extra computational time. An understanding of realistic levels of potential barrier fluctuations would be necessary for this to give deeper insight to our results. The use of a single-valued potential-barrier height here is equivalent to an assumption that fluctuations in thickness contribute more to IVs than fluctuations in barrier height.

We can complement the information shown from the fitted IV curves with measurements of the breakdown voltage, the point indicated by a star in Fig.~\ref{fig:dc_meas}~(b). We present a histogram showing the analysis of junction breakdown voltages of the same group of junctions in Fig.~\ref{fig:breakdown_stats}~(a). We phenomenologically fit a double-Gaussian to the breakdown voltage showing a bimodal distribution with a midpoint of 1.3 V. 
Breakdown occurs at the weakest point in the barrier at a voltage given by $V_{\rm BD} = tE_{\rm ds}$ where t is the barrier thickness and $E_{\rm ds}$ its dielectric strength. 
Assuming a uniform dielectric strength across the barrier, the weakest point is determined by the thinnest point in the junction. 
Dielectric strength could also fluctuate across the barrier, for instance decreasing at defects in the barrier, where what constitutes a defect in an amorphous oxide is a complicated issue in itself~\cite{strand2024structure}. If all junctions have the same thickness at their thinnest point, then the spread in Fig.~\ref{fig:breakdown_stats}~(a) would be a direct measure of the spread in dielectric strength, allowing us to place a bound on the standard deviation of this material property in these barriers, $\sigma_{E_{\rm DS}} < 4$~\%.
We consider the case where dielectric strength is constant across all barriers and within each barrier, i.e. omitting any contributions from local structural changes. In this limit the breakdown voltage is determined by the thinnest point of each barrier and the histogram in Fig.~\ref{fig:breakdown_stats}~(a) is a rescaled measurement of these thinnest points from 597 junctions. 

Extending the Monte-Carlo simulations shown in Fig.~\ref{fig:mciv} we consider whether the same model of varying barrier thickness agrees with our experimental measurements of breakdown voltage. 
Using thickness distributions identified in Fig.~\ref{fig:mciv} as giving the matches between simulated and experimental IVs, we compute a distribution of breakdown voltages. For Normal (Lognormal) distributions we use a thickness standard deviation and average of 0.21 and 1.025~nm  (0.155 and 1~nm) respectively.

We generate thicknesses from the distributions within the region of experimental matching in Fig.~\ref{fig:mciv} for a $240\times240$~nm junction with a 0.2~nm pixel size~\footnote{0.2~nm approximates the ionic radii in the AlO$_{\rm x}$ barrier}. We record the minimum thickness of the barrier in the junction and then repeat this to build a distribution of the thinnest points in the barrier. 
In order to translate the distribution of the thinnest points to breakdown voltages we rescale the distribution of thin-points by a constant dielectric strength of 3.31~GV/m and 5.26~GV/m for log-normal and normal barrier distributions respectively. Previous measurements of $E_{ds}$ in amorphous alumina vary from 0.4 to 1.0~GV/m , similar in order of magnitude~\cite{kolodzey2000electrical}.
We compute the dielectric strength by enforcing the mean of the rescaled thicknesses to be equal to the mean of the measured breakdown voltages. 
Simulated breakdown-distributions for barriers with log-normally and normally distributed thickness are shown in Fig.~\ref{fig:breakdown_stats} (b) and (c) respectively. Neither statistical distribution recreates the bimodal distribution seen experimentally.
However, the rescaled values from log-normally distributed barriers (Fig.~\ref{fig:breakdown_stats}~b) show reasonably good agreement with the measured breakdown voltages with a similar standard deviation and min/max ratio. 
In contrast the rescaled values taken from barriers generated with a normally distributed thickness (Fig.~\ref{fig:breakdown_stats}~c) do not agree with the experimental measurements. 
We show that multiple experimental measurements can be largely recreated considering only a skewed thickness distribution in our barriers. The parameters used to simulate breakdown are drawn from simulations of IV curves which have no awareness of breakdown physics and only one value is used to rescale the distribution. We note we do not recreate a bimodal distribution  indicating that, while better than a normal distribution, the log-normal thickness distribution is not a perfect description of the barrier.

In Fig.~\ref{fig:breakdown_stats}~(d) we present histograms of resistance for junctions which have been grouped by the breakdown voltage of the junction, distinguishing junctions as being above or below the mid-point of the bimodal distribution, 1.3~V.  
We show a difference in average resistance of the high vs. low breakdown junctions of $\sim$300~$\Omega$, a $\sim$4\% change. 
 
Here we show how a fabrication process can be assessed considering breakdown voltages, where the distribution of breakdown voltages could be used as feedback when optimizing junction growth protocols and correlations between breakdown voltages and resistances can be used to identify whether variation between barriers are contributing significantly to resistance spread. Specifically, we can identify two sub-ensembles of junctions within the nominally identical family of junctions which cannot be resolved from resistance measurements alone.
As mentioned in the introduction, it has recently been suggested that these thinnest points in junctions can contribute deviations of the C$\phi$R from an ideal sinusoidal relationship, causing effects such as Josephson harmonics~\cite{willsch2024observation, bayros2024influence}. This provides additional motivation to use this type of optimisation protocol as it may allow manufacturing processes to realise more homogeneous C$\phi$Rs.

In Fig.~\ref{fig:breakdown_stats}~(e) we show the thickness distribution of two example simulated junctions. We find that the distributions overlap quite well for counts at thinner distributions. Given that the junctions have similar resistances this is good evidence that the majority of the conductance is determined by the thinner portions of the barrier. We see this again in Fig.~\ref{fig:breakdown_stats}~(f) where we plot a cumulative conductance distribution as a function of junction area and find a large contribution to total conductance coming from a small fraction of the total area, irrespective of whether using a skewed lognormal distribution or a normally distributed thickness. Areas of locally elevated conductance are more prominent in the normally distributed junction, as is to be expected. 
As the thicker regions contribute little to the total conductance, our model has bad resolution on the thicker-side of the thickness distribution. The key feature of the log-normal probability distribution that our modeling does provide confidence for, is that the tails on the thinner-side of the distribution are suppressed more strongly than in the normal distribution. 
The observation of a small fraction of the junction contributing a large fraction of conductance agrees qualitatively with the conclusions from Refs.~\cite{aref2014characterization, oh2025correlating}. Both studies arrive at this conclusion from considerations of barrier thickness measurements by (S)TEM, which we consider in the next section.

\section{Scanning Transmission Electron Microscopy}

\begin{figure}
    \centering
    \includegraphics[width=\linewidth]{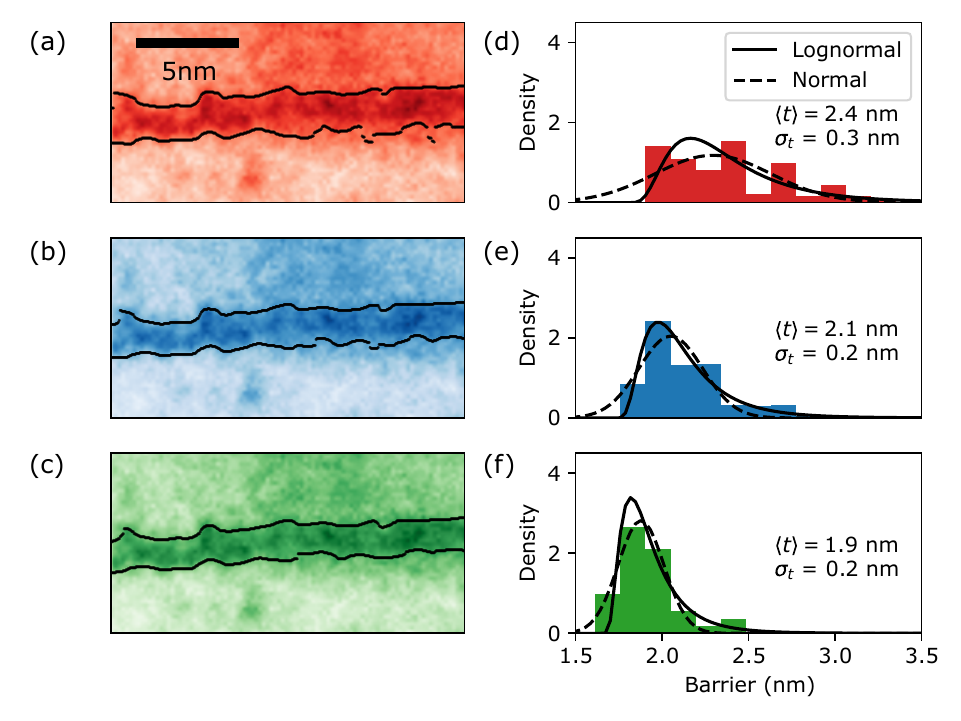}
    \caption{Experimental STEM-EDS imaging of an Al/AlO$_{\rm x}$/Al barrier of the oxygen peak. (a-c) show heatmaps indicating the oxygen peak with the edges of the barrier as detected by Kernel integration overlaid to the maps. Different values of $\delta$ are used to define the kernel for the three maps (0, 0.2, 0.4 respectively). (d-f) show histograms of thicknesses as inferred from the different kernel integrations. }
    \label{fig:STEM}
\end{figure}

STEM includes a family of techniques which allow for sub-nanometre resolution imaging of materials with elemental contrast which are natural candidates to investigate questions about the thickness distribution of the barrier. These techniques can be applied to JJs where the barrier has been capped with a top electrode, i.e. after the full fabrication process has taken place. 
We first consider a protocol to measure variations in thickness in the projected 2D image. We do this considering images of the oxygen content in a barrier collected by STEM energy-dispersive X-ray spectroscopy (EDS) mapping. Imaging is performed on a lamella which is fabricated by focused ion beam cutting and lift out of a $\sim$30~nm thick cross section of the JJ. In Fig.~\ref{fig:STEM} we present an example image of the oxygen composition. The edge of the barrier is identified by kernel integration using OpenCV~\cite{opencv_library}. 
We define two kernels to identify the two edges where each kernel has an asymmetry factor $\delta$, which differentially changes the weight of pixels on the left and right hand side of the kernel. We describe the construction of the integration kernels in the supplemental information. 
Given that the oxygen content in the barrier increases over a finite length, the edge of the barrier could be defined at several points. By changing $\delta$ we systematically alter where the kernel integration protocol identifies the edge. 
In Fig.~\ref{fig:STEM}~(a-c) we show the same image with edges identified using 0, 0.2, 0.4 as values of $\delta$ respectively. In analysis we explored larger values of $\delta$ and found the edges were obviously within the junction, so care must be taken in selecting this asymmetry value. 

We measure the thickness of the barrier by finding the distance between the two edges for each column of pixels. The average of the thickness from these different kernels matches previous measurements by TEM well giving an average thickness of $\sim$~1.9~nm - 2.4~nm depending on the kernel~\cite{zeng2015direct}. 
We can also compute the distribution of thicknesses shown in the histograms Fig.~\ref{fig:STEM}~(d-f) and fit each of them both to normal and log-normal distributions. We routinely see that the log-normal distribution fits the thickness distribution better than the normal distribution. In the literature the results of this type of analysis have been interpreted as the statistical distribution of thicknesses of a barrier, a claim we are explicitly not making here. 

Arbitrary choices, such as the asymmetry factor in the kernels, can change what we infer about the barrier. We attribute this both to (i) details of the barrier including such as the oxygen content likely changing over a finite distance and the non-uniform barrier thickness and (ii) the STEM image being a projection which averages the barrier. As the barrier is grown on a bottom electrode which isn't perfectly flat the imaging beam is at times intersecting both electrode and barrier.
We wish to understand (i) from techniques such as STEM imaging as they will determine the behaviour of a JJ. However (ii) represents difficulties posed in trying to identify thickness variations in this material system by STEM.
We suggest that we should report 1.9~nm $< \langle t \rangle <$ 2.4~nm, i.e. the variation in thickness implied by different kernels is what defines the error bars on the measurement of average barrier thickness.

In order to gain a deeper understanding of (ii) from above, we simulate STEM EDS images of oxygen content.
We simulate the STEM image by building a 3D array of 0.1~nm$^3$ voxels which represents the lamella used for STEM imaging. We use AFM maps to determine the topography of the barrier and then assign voxels in the barrier a value of 1, representing high oxygen content, voxels on the edge of the barrier a value of 0.5 and other voxels a value of 0, representing low oxygen content. 
Averaging values of voxels along a column normal to the lamella surface gives a value corresponding to the nominal oxygen content in that column and represents the projection formed by the electron beam experimentally. 
After projection we add Gaussian noise to the image representative of detector noise, we calibrate the centre and standard deviation of the Gaussian noise from a line trace of the experimental data in Fig.~\ref{fig:STEM}. We finally apply a Gaussian blur with radius 0.1~nm representative of the beam size in STEM experiments.

\begin{figure}
    \centering
    \includegraphics[width=\linewidth]{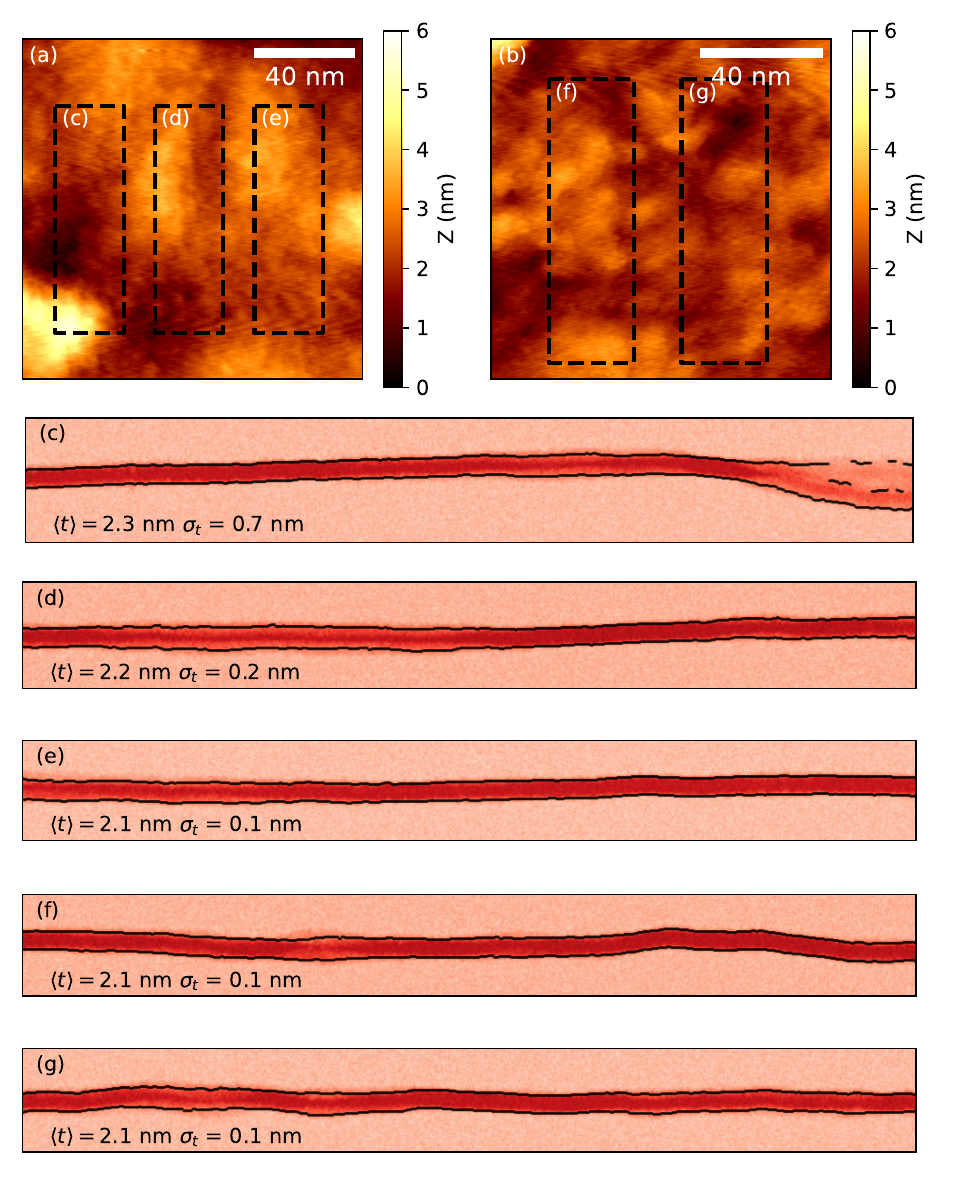}
    \caption{(a, b) two high resolution AFM scans of representative bottom leads (region indicated in Fig.~\ref{fig:dc_meas}~a). The route mean squared roughness for the areas shown are 0.7 and 0.5~nm respectively. In (a, b) five 30~$\times$~100~nm areas are marked by dashed lines. These regions are used to simulate STEM profiles shown in (c-g) assuming a 2~nm oxide barrier coating the AFM topography with the labels matching the AFM region to the STEM profile inserted into the boxes. (c-g) Edges of simulated oxygen content are detected using a symmetric kernels and shown overlaid, as well as measurements of average thickness and thickness standard deviations being noted in each figure. The labels (c - g) are in the same side of the scan region in the simulated STEM profile and the AFM regions. }
    \label{fig:afm_stem}
\end{figure}

We generate a series of barriers assuming that the thickness of the barrier is uniformly 2~nm across the junction. Using a uniform barrier thickness means that any thickness variation inferred from the simulated STEM image is fictitious, indicative of what we may expect to see from experimental nonidealities which could change what we infer from analysis of STEM images.  
We draw the bottom electrode topography from AFM measurements of the bottom lead shown in Fig.~\ref{fig:afm_stem}~(a,b). These scans were taken from a region approximately overlapping the text `Bottom Lead' in Fig.~\ref{fig:dc_meas}~(a). \footnote{Noise at the $\sim$0.6~nm range is present in some regions of these AFM scans, which we found hard to exclude from measurements. This represents $\sim$10~\% of the total topographic variation in these AFM maps and is not present across the whole scans and therefore we believe not a dominant contributor to the following analysis.}. 
We infer the thickness of the barriers from the simulated STEM images (Fig.~\ref{fig:afm_stem} c-g) using symmetric kernels ($\delta = 0$) as described for Fig.~\ref{fig:STEM} adjusting kernel sizes to account for the different pixel sizes. 
As expected, a perfectly uniform barrier appears to have varying thickness when viewed in projection due to roughness in the underlying barrier. 
We also find that the thickness values and thickness standard deviation change depending which AFM area is used to build the projection. All but one average thickness measurements are within 10~\% of the barrier thickness input to the simulation, although the standard deviation of barrier thickness varies by a factor of $\sim7\times$. 
This suggests that STEM is a good technique to measure the average thickness of a barrier (reporting errors as described above) but calls into question the robustness of using STEM profiles to infer the thickness distribution in a junction. 
It also shows that multiple areas should be measured and inferring average thickness values from single STEM images may introduce errors. We therefore include additional STEM images with analogous processing to that performed in Fig.~\ref{fig:STEM} in the supplemental material. These suggest similar thicknesses, but also include regions of topographic features which slightly increase the inferred average thickness.

Sometimes it is straightforward to identify regions where apparent thickness broadening occurs due to topographic variation in the bottom electrode (e.g. RHS of Fig.~\ref{fig:afm_stem}~c) where it's possible to resolve two barriers superimposed on one another. Here the kernel edge detection does not work well, shown by the discontinuities in the edge it identifies. In the supplemental material we show an example of experimental STEM imaging with these types of features, compared to a similar STEM simulation. 
In Fig.~\ref{fig:afm_stem}~(d) there is also a topographic gradient along the projection axis, but, in this instance, the topographic change is smaller in magnitude and so in the simulated STEM image it appears only as a broadening of the barrier. 
It is therefore not straightforwardly possible to systematically distinguish apparent broadening from topography variation and true barrier variation, although it is possible to exclude areas where the topographic broadening effects are largest.

Comparing experimental (Fig.~\ref{fig:STEM}) and simulated (Fig.~\ref{fig:afm_stem}) STEM images, barriers appear rougher in the experimental STEM data than the simulated STEM. 
The topography of the electrode used to simulate STEM images may be artificially smoothed due to the AFM tip radius of 2~nm. This would mean that features of the size of apparent roughness seen in Fig.~\ref{fig:STEM} would not be apparent with this AFM tip. The barriers we measure here are also oxidised which can change the topography of the surface. 
While there may be features sufficiently small that they are not resolved by AFM, both AFM maps in Fig.~\ref{fig:afm_stem} have topographic features that are large relative to the AFM resolution, and as we show, couple importantly into the inferred STEM projections. Using STM may allow higher resolution topographic maps to perform more robust inferences of STEM profiles, and can also be performed \emph{in situ} before barrier oxidation modifies the topography. These more detailed maps might find uses, for instance in reverse Monte-Carlo approaches~\cite{mcgreevy2001reverse} to extract quantitative information about the barrier thickness and thickness distributions from STEM. 

\section{Conclusions}

\begin{table*}
\centering
\renewcommand{\arraystretch}{1.2}
\begin{tabular}{|c|c|cccc|ccc|}
\hline
 \textbf{Technique} & \begin{tabular}[c]{@{}c@{}}\textbf{Simmons} \\\textbf{Model Fit}\end{tabular} & \multicolumn{4}{c|}{\begin{tabular}[c]{@{}c@{}c@{}}\textbf{Monte-Carlo}\\\textbf{Simulations}\end{tabular}}                            & \multicolumn{3}{c|}{\textbf{STEM-EDS}}                            \\ \hline
 \multirow{2}{*}{\textbf{Variable}} &  \multirow{2}{*}{-} & \multicolumn{4}{c|}{\begin{tabular}[c]{@{}c@{}c@{}}\textbf{Potential Barrier Height (eV)}\\ \textbf{Only log-normal distribution}\end{tabular}}                            & \multicolumn{3}{c|}{\textbf{Kernel Asymmetry ($\delta$)}}                            \\ \cline{3-9} 
  &   & \multicolumn{1}{c|}{0.8} & \multicolumn{1}{c|}{1.0} & \multicolumn{1}{c|}{1.22} & 1.5 & \multicolumn{1}{c|}{0} & \multicolumn{1}{c|}{0.2} &  0.4\\ \hline
 \begin{tabular}[c]{@{}c@{}}\textbf{Thickness}\\\textbf{(nm)}\end{tabular}& 0.78~$\pm$~0.02 & \multicolumn{1}{c|}{No fit} & \multicolumn{1}{c|}{1.3$\pm$~0.3} &  \multicolumn{1}{c|}{1.0~$\pm$~0.2} & 0.8~$\pm$~0.1 & \multicolumn{1}{c|}{2.4~$\pm$~0.3} & \multicolumn{1}{c|}{2.1~$\pm$~0.2} &  1.9~$\pm$~0.2\\ \hline
\end{tabular}

\caption{A summary of different barrier thicknesses extracted throughout this manuscript. 
Fitting the Simmons model returns a single-valued thickness and the error represents the standard deviation from fitting 597 IVs. 
Monte-Carlo simulations give thickness by fixing the area of a junction and allowing the barrier thickness to vary. The thickness value is the average thickness whereas the errors represent a genuine thickness distribution in the barrier. 
STEM-EDS measurements are direct measurements of thickness by atomic scale microscopy. As shown here, the thickness of the barrier is sensitive to image processing (i.e. choice of $\delta$). Errors here represent standard deviations of inferred average thickness for a specific choice of $\delta$ but do not represent thickness distributions. Without an un-biased approach to image processing, the true error on thickness measurements is likely the variation in thickness between different values of $\delta$.}
\label{tab:summary}
\end{table*}

In this article we have aimed to understand oxide barriers in JJs due to their technological importance in superconducting qubits. A summary of different thicknesses inferred within this paper is shown in Table~\ref{tab:summary}. 
We fit the Simmons model to non-linear IVs to extract a single-valued thickness value to each junction $\sim$0.78~nm. 
We build Monte-Carlo protocols to simulate the IV properties of barriers with multi-valued thicknesses drawn from example statistical distributions. Here we find that the distribution of thicknesses present in the junction, results in IVs that, when fitted, would imply a single-value thickness smaller than the average thickness. This is because conductivity is exponentially dependent on the barrier thickness, so any thinner-than-average regions contribute more than similarly thicker-than-average regions. These simulations indicate a barrier that is $\sim$1~nm thick with a standard deviation of $\sim$0.2~nm, however using different barrier heights, we infer different thickness distributions. The rectangular potential barrier used in the Simmons model likely results in an under-estimate of thickness relative to more realistic, smoothly increasing barriers. By considering the Monte Carlo simulations relative to STEM measurements we constrain the barrier height to the range 0.8~eV $\lesssim \phi \lesssim $~1.5~eV with best matching at $\sim$~1.2~eV.

Using breakdown measurements we can probe the tails of thickness distributions assuming that breakdown occurs at the thinnest point or points of the barrier. Considering normal and skewed log-normal distributions of barrier thickness, we show that the breakdown statistics are better described by a skewed distribution. 
We show that the tails of the thickness distribution contribute substantially to the resistance of the device with a $\sim$4~\% difference in resistance between the 93~\% lowest- and 7~\% highest-breakdown voltage junctions. We also show that normally distributed barrier thicknesses give rise to larger barrier-induced spread than skewed distributions.

We then consider STEM measurements probing the oxygen content of barriers as a technique to directly measure the thickness of barriers. We show measurements of the barrier thickness from 1.9 to 2.4~nm depending on how the edge of the barrier is determined with a standard deviation not dissimilar to those inferred from the Monte Carlo simulations of the barrier. 
We suggest an approach to define errors on barrier thickness measurements, performing kernel integration with different kernels to extract different barrier edges and comparing the variation in the average thickness reported. These error bars should not be misinterpreted as the standard deviation of the barrier thickness. 

Comparing experimental measurements to simulations of STEM data, we show that inferring average thicknesses from STEM images is reasonable, subject to errors associated with identifying the edge of the barrier. However, inferring variations in the thickness of the barrier from STEM images is not a robust process. The roughness commonly seen on bottom electrodes, when projected into 2D, also creates apparent thickness variations. 
We see wider variations in barrier thickness measurements experimentally than when simulating a perfectly uniform barrier, lending support to the idea that the barrier thickness does indeed vary across the junction but inferring the statistics of this variation from STEM data alone remains a challenge. 
Most evidence in the literature about thickness distributions of barriers and that this is normally distributed arises from this type of processing of STEM or TEM data for instance~\cite{zeng2015direct}. While these works were substantial increments to the understanding of the oxide barriers in JJs, continuing to use these techniques and inferences should be avoided, for instance in computations of Josephson harmonics based on normal thickness distributions~\cite{willsch2024observation}. We show here that this type of inference over-states the certainty of thickness distributions, and indeed our most sensitive probe of the tails of our distribution imply some form of skewed thickness distribution contradicting the conclusions drawn from STEM.

\begin{acknowledgments}
The authors thank the OQC Fabrication team for discussions and contributions toward our Josephson junction manufacturing process. 
The STEM imaging was taken in LEMAS center in the University of Leeds, thanks to John Harrington and Zabeada Aslam for FIB sectioning and STEM operation.
We thank the Royal Holloway University of London SuperFab Facility for their support.
\end{acknowledgments}

\bibliography{bibliography}

\begin{thebibliography}{47}%
\makeatletter
\providecommand \@ifxundefined [1]{%
 \@ifx{#1\undefined}
}%
\providecommand \@ifnum [1]{%
 \ifnum #1\expandafter \@firstoftwo
 \else \expandafter \@secondoftwo
 \fi
}%
\providecommand \@ifx [1]{%
 \ifx #1\expandafter \@firstoftwo
 \else \expandafter \@secondoftwo
 \fi
}%
\providecommand \natexlab [1]{#1}%
\providecommand \enquote  [1]{``#1''}%
\providecommand \bibnamefont  [1]{#1}%
\providecommand \bibfnamefont [1]{#1}%
\providecommand \citenamefont [1]{#1}%
\providecommand \href@noop [0]{\@secondoftwo}%
\providecommand \href [0]{\begingroup \@sanitize@url \@href}%
\providecommand \@href[1]{\@@startlink{#1}\@@href}%
\providecommand \@@href[1]{\endgroup#1\@@endlink}%
\providecommand \@sanitize@url [0]{\catcode `\\12\catcode `\$12\catcode `\&12\catcode `\#12\catcode `\^12\catcode `\_12\catcode `\%12\relax}%
\providecommand \@@startlink[1]{}%
\providecommand \@@endlink[0]{}%
\providecommand \url  [0]{\begingroup\@sanitize@url \@url }%
\providecommand \@url [1]{\endgroup\@href {#1}{\urlprefix }}%
\providecommand \urlprefix  [0]{URL }%
\providecommand \Eprint [0]{\href }%
\providecommand \doibase [0]{https://doi.org/}%
\providecommand \selectlanguage [0]{\@gobble}%
\providecommand \bibinfo  [0]{\@secondoftwo}%
\providecommand \bibfield  [0]{\@secondoftwo}%
\providecommand \translation [1]{[#1]}%
\providecommand \BibitemOpen [0]{}%
\providecommand \bibitemStop [0]{}%
\providecommand \bibitemNoStop [0]{.\EOS\space}%
\providecommand \EOS [0]{\spacefactor3000\relax}%
\providecommand \BibitemShut  [1]{\csname bibitem#1\endcsname}%
\let\auto@bib@innerbib\@empty
\bibitem [{\citenamefont {Arute}\ \emph {et~al.}(2019)\citenamefont {Arute}, \citenamefont {Arya}, \citenamefont {Babbush}, \citenamefont {Bacon}, \citenamefont {Bardin}, \citenamefont {Barends}, \citenamefont {Biswas}, \citenamefont {Boixo}, \citenamefont {Brandao}, \citenamefont {Buell} \emph {et~al.}}]{arute2019quantum}%
  \BibitemOpen
  \bibfield  {author} {\bibinfo {author} {\bibfnamefont {F.}~\bibnamefont {Arute}}, \bibinfo {author} {\bibfnamefont {K.}~\bibnamefont {Arya}}, \bibinfo {author} {\bibfnamefont {R.}~\bibnamefont {Babbush}}, \bibinfo {author} {\bibfnamefont {D.}~\bibnamefont {Bacon}}, \bibinfo {author} {\bibfnamefont {J.~C.}\ \bibnamefont {Bardin}}, \bibinfo {author} {\bibfnamefont {R.}~\bibnamefont {Barends}}, \bibinfo {author} {\bibfnamefont {R.}~\bibnamefont {Biswas}}, \bibinfo {author} {\bibfnamefont {S.}~\bibnamefont {Boixo}}, \bibinfo {author} {\bibfnamefont {F.~G.}\ \bibnamefont {Brandao}}, \bibinfo {author} {\bibfnamefont {D.~A.}\ \bibnamefont {Buell}}, \emph {et~al.},\ }\bibfield  {title} {\bibinfo {title} {Quantum supremacy using a programmable superconducting processor},\ }\href@noop {} {\bibfield  {journal} {\bibinfo  {journal} {Nature}\ }\textbf {\bibinfo {volume} {574}},\ \bibinfo {pages} {505} (\bibinfo {year} {2019})}\BibitemShut {NoStop}%
\bibitem [{\citenamefont {Macklin}\ \emph {et~al.}(2015)\citenamefont {Macklin}, \citenamefont {O’brien}, \citenamefont {Hover}, \citenamefont {Schwartz}, \citenamefont {Bolkhovsky}, \citenamefont {Zhang}, \citenamefont {Oliver},\ and\ \citenamefont {Siddiqi}}]{macklin2015near}%
  \BibitemOpen
  \bibfield  {author} {\bibinfo {author} {\bibfnamefont {C.}~\bibnamefont {Macklin}}, \bibinfo {author} {\bibfnamefont {K.}~\bibnamefont {O’brien}}, \bibinfo {author} {\bibfnamefont {D.}~\bibnamefont {Hover}}, \bibinfo {author} {\bibfnamefont {M.}~\bibnamefont {Schwartz}}, \bibinfo {author} {\bibfnamefont {V.}~\bibnamefont {Bolkhovsky}}, \bibinfo {author} {\bibfnamefont {X.}~\bibnamefont {Zhang}}, \bibinfo {author} {\bibfnamefont {W.}~\bibnamefont {Oliver}},\ and\ \bibinfo {author} {\bibfnamefont {I.}~\bibnamefont {Siddiqi}},\ }\bibfield  {title} {\bibinfo {title} {A near--quantum-limited josephson traveling-wave parametric amplifier},\ }\href@noop {} {\bibfield  {journal} {\bibinfo  {journal} {Science}\ }\textbf {\bibinfo {volume} {350}},\ \bibinfo {pages} {307} (\bibinfo {year} {2015})}\BibitemShut {NoStop}%
\bibitem [{\citenamefont {Nik}\ \emph {et~al.}(2016)\citenamefont {Nik}, \citenamefont {Krantz}, \citenamefont {Zeng}, \citenamefont {Greibe}, \citenamefont {Pettersson}, \citenamefont {Gustafsson}, \citenamefont {Delsing},\ and\ \citenamefont {Olsson}}]{nik2016correlation}%
  \BibitemOpen
  \bibfield  {author} {\bibinfo {author} {\bibfnamefont {S.}~\bibnamefont {Nik}}, \bibinfo {author} {\bibfnamefont {P.}~\bibnamefont {Krantz}}, \bibinfo {author} {\bibfnamefont {L.}~\bibnamefont {Zeng}}, \bibinfo {author} {\bibfnamefont {T.}~\bibnamefont {Greibe}}, \bibinfo {author} {\bibfnamefont {H.}~\bibnamefont {Pettersson}}, \bibinfo {author} {\bibfnamefont {S.}~\bibnamefont {Gustafsson}}, \bibinfo {author} {\bibfnamefont {P.}~\bibnamefont {Delsing}},\ and\ \bibinfo {author} {\bibfnamefont {E.}~\bibnamefont {Olsson}},\ }\bibfield  {title} {\bibinfo {title} {Correlation between al grain size, grain boundary grooves and local variations in oxide barrier thickness of al/alo x/al tunnel junctions by transmission electron microscopy},\ }\href@noop {} {\bibfield  {journal} {\bibinfo  {journal} {SpringerPlus}\ }\textbf {\bibinfo {volume} {5}},\ \bibinfo {pages} {1} (\bibinfo {year} {2016})}\BibitemShut {NoStop}%
\bibitem [{\citenamefont {Fritz}\ \emph {et~al.}(2019{\natexlab{a}})\citenamefont {Fritz}, \citenamefont {Radtke}, \citenamefont {Schneider}, \citenamefont {Luysberg}, \citenamefont {Weides},\ and\ \citenamefont {Gerthsen}}]{fritz2019structural}%
  \BibitemOpen
  \bibfield  {author} {\bibinfo {author} {\bibfnamefont {S.}~\bibnamefont {Fritz}}, \bibinfo {author} {\bibfnamefont {L.}~\bibnamefont {Radtke}}, \bibinfo {author} {\bibfnamefont {R.}~\bibnamefont {Schneider}}, \bibinfo {author} {\bibfnamefont {M.}~\bibnamefont {Luysberg}}, \bibinfo {author} {\bibfnamefont {M.}~\bibnamefont {Weides}},\ and\ \bibinfo {author} {\bibfnamefont {D.}~\bibnamefont {Gerthsen}},\ }\bibfield  {title} {\bibinfo {title} {Structural and nanochemical properties of al o x layers in al/al o x/al-layer systems for josephson junctions},\ }\href@noop {} {\bibfield  {journal} {\bibinfo  {journal} {Physical Review Materials}\ }\textbf {\bibinfo {volume} {3}},\ \bibinfo {pages} {114805} (\bibinfo {year} {2019}{\natexlab{a}})}\BibitemShut {NoStop}%
\bibitem [{\citenamefont {Zeng}\ \emph {et~al.}(2016)\citenamefont {Zeng}, \citenamefont {Tran}, \citenamefont {Tai}, \citenamefont {Svensson},\ and\ \citenamefont {Olsson}}]{zeng2016atomic}%
  \BibitemOpen
  \bibfield  {author} {\bibinfo {author} {\bibfnamefont {L.}~\bibnamefont {Zeng}}, \bibinfo {author} {\bibfnamefont {D.~T.}\ \bibnamefont {Tran}}, \bibinfo {author} {\bibfnamefont {C.-W.}\ \bibnamefont {Tai}}, \bibinfo {author} {\bibfnamefont {G.}~\bibnamefont {Svensson}},\ and\ \bibinfo {author} {\bibfnamefont {E.}~\bibnamefont {Olsson}},\ }\bibfield  {title} {\bibinfo {title} {Atomic structure and oxygen deficiency of the ultrathin aluminium oxide barrier in al/alox/al josephson junctions},\ }\href@noop {} {\bibfield  {journal} {\bibinfo  {journal} {Scientific reports}\ }\textbf {\bibinfo {volume} {6}},\ \bibinfo {pages} {29679} (\bibinfo {year} {2016})}\BibitemShut {NoStop}%
\bibitem [{\citenamefont {Lapham}\ and\ \citenamefont {Georgiev}(2022)}]{lapham2022computational}%
  \BibitemOpen
  \bibfield  {author} {\bibinfo {author} {\bibfnamefont {P.}~\bibnamefont {Lapham}}\ and\ \bibinfo {author} {\bibfnamefont {V.~P.}\ \bibnamefont {Georgiev}},\ }\bibfield  {title} {\bibinfo {title} {Computational study of oxide stoichiometry and variability in the al/alox/al tunnel junction},\ }\href@noop {} {\bibfield  {journal} {\bibinfo  {journal} {Nanotechnology}\ }\textbf {\bibinfo {volume} {33}},\ \bibinfo {pages} {265201} (\bibinfo {year} {2022})}\BibitemShut {NoStop}%
\bibitem [{\citenamefont {Yohannes}\ \emph {et~al.}(2005)\citenamefont {Yohannes}, \citenamefont {Sarwana}, \citenamefont {Tolpygo}, \citenamefont {Sahu}, \citenamefont {Polyakov},\ and\ \citenamefont {Semenov}}]{yohannes2005characterization}%
  \BibitemOpen
  \bibfield  {author} {\bibinfo {author} {\bibfnamefont {D.}~\bibnamefont {Yohannes}}, \bibinfo {author} {\bibfnamefont {S.}~\bibnamefont {Sarwana}}, \bibinfo {author} {\bibfnamefont {S.~K.}\ \bibnamefont {Tolpygo}}, \bibinfo {author} {\bibfnamefont {A.}~\bibnamefont {Sahu}}, \bibinfo {author} {\bibfnamefont {Y.~A.}\ \bibnamefont {Polyakov}},\ and\ \bibinfo {author} {\bibfnamefont {V.~K.}\ \bibnamefont {Semenov}},\ }\bibfield  {title} {\bibinfo {title} {Characterization of hypres'4.5 ka/cm/sup 2/\& 8 ka/cm/sup 2/nb/alo/sub x//nb fabrication processes},\ }\href@noop {} {\bibfield  {journal} {\bibinfo  {journal} {IEEE Transactions on Applied Superconductivity}\ }\textbf {\bibinfo {volume} {15}},\ \bibinfo {pages} {90} (\bibinfo {year} {2005})}\BibitemShut {NoStop}%
\bibitem [{\citenamefont {West}\ \emph {et~al.}(2022)\citenamefont {West}, \citenamefont {Kurlej}, \citenamefont {Wynn}, \citenamefont {Rogers}, \citenamefont {Gouker},\ and\ \citenamefont {Tolpygo}}]{west2022wafer}%
  \BibitemOpen
  \bibfield  {author} {\bibinfo {author} {\bibfnamefont {J.~T.}\ \bibnamefont {West}}, \bibinfo {author} {\bibfnamefont {A.}~\bibnamefont {Kurlej}}, \bibinfo {author} {\bibfnamefont {A.}~\bibnamefont {Wynn}}, \bibinfo {author} {\bibfnamefont {C.}~\bibnamefont {Rogers}}, \bibinfo {author} {\bibfnamefont {M.~A.}\ \bibnamefont {Gouker}},\ and\ \bibinfo {author} {\bibfnamefont {S.~K.}\ \bibnamefont {Tolpygo}},\ }\bibfield  {title} {\bibinfo {title} {Wafer-scale characterization of a superconductor integrated circuit fabrication process, using a cryogenic wafer prober},\ }\href@noop {} {\bibfield  {journal} {\bibinfo  {journal} {IEEE Transactions on Applied Superconductivity}\ }\textbf {\bibinfo {volume} {32}},\ \bibinfo {pages} {1} (\bibinfo {year} {2022})}\BibitemShut {NoStop}%
\bibitem [{\citenamefont {Semenov}\ \emph {et~al.}(2019)\citenamefont {Semenov}, \citenamefont {Polyakov},\ and\ \citenamefont {Tolpygo}}]{semenov2019very}%
  \BibitemOpen
  \bibfield  {author} {\bibinfo {author} {\bibfnamefont {V.~K.}\ \bibnamefont {Semenov}}, \bibinfo {author} {\bibfnamefont {Y.~A.}\ \bibnamefont {Polyakov}},\ and\ \bibinfo {author} {\bibfnamefont {S.~K.}\ \bibnamefont {Tolpygo}},\ }\bibfield  {title} {\bibinfo {title} {Very large scale integration of josephson-junction-based superconductor random access memories},\ }\href@noop {} {\bibfield  {journal} {\bibinfo  {journal} {IEEE Transactions on Applied Superconductivity}\ }\textbf {\bibinfo {volume} {29}},\ \bibinfo {pages} {1} (\bibinfo {year} {2019})}\BibitemShut {NoStop}%
\bibitem [{\citenamefont {Kreikebaum}\ \emph {et~al.}(2020)\citenamefont {Kreikebaum}, \citenamefont {O’Brien}, \citenamefont {Morvan},\ and\ \citenamefont {Siddiqi}}]{kreikebaum2020improving}%
  \BibitemOpen
  \bibfield  {author} {\bibinfo {author} {\bibfnamefont {J.}~\bibnamefont {Kreikebaum}}, \bibinfo {author} {\bibfnamefont {K.}~\bibnamefont {O’Brien}}, \bibinfo {author} {\bibfnamefont {A.}~\bibnamefont {Morvan}},\ and\ \bibinfo {author} {\bibfnamefont {I.}~\bibnamefont {Siddiqi}},\ }\bibfield  {title} {\bibinfo {title} {Improving wafer-scale josephson junction resistance variation in superconducting quantum coherent circuits},\ }\href@noop {} {\bibfield  {journal} {\bibinfo  {journal} {Superconductor Science and Technology}\ }\textbf {\bibinfo {volume} {33}},\ \bibinfo {pages} {06LT02} (\bibinfo {year} {2020})}\BibitemShut {NoStop}%
\bibitem [{\citenamefont {Ambegaokar}\ and\ \citenamefont {Baratoff}(1963)}]{ambegaokar1963tunneling}%
  \BibitemOpen
  \bibfield  {author} {\bibinfo {author} {\bibfnamefont {V.}~\bibnamefont {Ambegaokar}}\ and\ \bibinfo {author} {\bibfnamefont {A.}~\bibnamefont {Baratoff}},\ }\bibfield  {title} {\bibinfo {title} {Tunneling between superconductors},\ }\href@noop {} {\bibfield  {journal} {\bibinfo  {journal} {Physical review letters}\ }\textbf {\bibinfo {volume} {10}},\ \bibinfo {pages} {486} (\bibinfo {year} {1963})}\BibitemShut {NoStop}%
\bibitem [{\citenamefont {Osman}\ \emph {et~al.}(2021)\citenamefont {Osman}, \citenamefont {Simon}, \citenamefont {Bengtsson}, \citenamefont {Kosen}, \citenamefont {Krantz}, \citenamefont {P~Lozano}, \citenamefont {Scigliuzzo}, \citenamefont {Delsing}, \citenamefont {Bylander},\ and\ \citenamefont {Fadavi~Roudsari}}]{osman2021simplified}%
  \BibitemOpen
  \bibfield  {author} {\bibinfo {author} {\bibfnamefont {A.}~\bibnamefont {Osman}}, \bibinfo {author} {\bibfnamefont {J.}~\bibnamefont {Simon}}, \bibinfo {author} {\bibfnamefont {A.}~\bibnamefont {Bengtsson}}, \bibinfo {author} {\bibfnamefont {S.}~\bibnamefont {Kosen}}, \bibinfo {author} {\bibfnamefont {P.}~\bibnamefont {Krantz}}, \bibinfo {author} {\bibfnamefont {D.}~\bibnamefont {P~Lozano}}, \bibinfo {author} {\bibfnamefont {M.}~\bibnamefont {Scigliuzzo}}, \bibinfo {author} {\bibfnamefont {P.}~\bibnamefont {Delsing}}, \bibinfo {author} {\bibfnamefont {J.}~\bibnamefont {Bylander}},\ and\ \bibinfo {author} {\bibfnamefont {A.}~\bibnamefont {Fadavi~Roudsari}},\ }\bibfield  {title} {\bibinfo {title} {Simplified josephson-junction fabrication process for reproducibly high-performance superconducting qubits},\ }\href@noop {} {\bibfield  {journal} {\bibinfo  {journal} {Applied Physics Letters}\ }\textbf {\bibinfo {volume} {118}} (\bibinfo {year} {2021})}\BibitemShut {NoStop}%
\bibitem [{\citenamefont {Moskalev}\ \emph {et~al.}(2023)\citenamefont {Moskalev}, \citenamefont {Zikiy}, \citenamefont {Pishchimova}, \citenamefont {Ezenkova}, \citenamefont {Smirnov}, \citenamefont {Ivanov}, \citenamefont {Korshakov},\ and\ \citenamefont {Rodionov}}]{moskalev2023optimization}%
  \BibitemOpen
  \bibfield  {author} {\bibinfo {author} {\bibfnamefont {D.~O.}\ \bibnamefont {Moskalev}}, \bibinfo {author} {\bibfnamefont {E.~V.}\ \bibnamefont {Zikiy}}, \bibinfo {author} {\bibfnamefont {A.~A.}\ \bibnamefont {Pishchimova}}, \bibinfo {author} {\bibfnamefont {D.~A.}\ \bibnamefont {Ezenkova}}, \bibinfo {author} {\bibfnamefont {N.~S.}\ \bibnamefont {Smirnov}}, \bibinfo {author} {\bibfnamefont {A.~I.}\ \bibnamefont {Ivanov}}, \bibinfo {author} {\bibfnamefont {N.~D.}\ \bibnamefont {Korshakov}},\ and\ \bibinfo {author} {\bibfnamefont {I.~A.}\ \bibnamefont {Rodionov}},\ }\bibfield  {title} {\bibinfo {title} {Optimization of shadow evaporation and oxidation for reproducible quantum josephson junction circuits},\ }\href@noop {} {\bibfield  {journal} {\bibinfo  {journal} {Scientific Reports}\ }\textbf {\bibinfo {volume} {13}},\ \bibinfo {pages} {4174} (\bibinfo {year} {2023})}\BibitemShut {NoStop}%
\bibitem [{\citenamefont {Pishchimova}\ \emph {et~al.}(2023)\citenamefont {Pishchimova}, \citenamefont {Smirnov}, \citenamefont {Ezenkova}, \citenamefont {Krivko}, \citenamefont {Zikiy}, \citenamefont {Moskalev}, \citenamefont {Ivanov}, \citenamefont {Korshakov},\ and\ \citenamefont {Rodionov}}]{pishchimova2023improving}%
  \BibitemOpen
  \bibfield  {author} {\bibinfo {author} {\bibfnamefont {A.~A.}\ \bibnamefont {Pishchimova}}, \bibinfo {author} {\bibfnamefont {N.~S.}\ \bibnamefont {Smirnov}}, \bibinfo {author} {\bibfnamefont {D.~A.}\ \bibnamefont {Ezenkova}}, \bibinfo {author} {\bibfnamefont {E.~A.}\ \bibnamefont {Krivko}}, \bibinfo {author} {\bibfnamefont {E.~V.}\ \bibnamefont {Zikiy}}, \bibinfo {author} {\bibfnamefont {D.~O.}\ \bibnamefont {Moskalev}}, \bibinfo {author} {\bibfnamefont {A.~I.}\ \bibnamefont {Ivanov}}, \bibinfo {author} {\bibfnamefont {N.~D.}\ \bibnamefont {Korshakov}},\ and\ \bibinfo {author} {\bibfnamefont {I.~A.}\ \bibnamefont {Rodionov}},\ }\bibfield  {title} {\bibinfo {title} {Improving josephson junction reproducibility for superconducting quantum circuits: Junction area fluctuation},\ }\href@noop {} {\bibfield  {journal} {\bibinfo  {journal} {Scientific Reports}\ }\textbf {\bibinfo {volume} {13}},\ \bibinfo {pages} {6772} (\bibinfo {year} {2023})}\BibitemShut {NoStop}%
\bibitem [{\citenamefont {Osman}\ \emph {et~al.}(2023)\citenamefont {Osman}, \citenamefont {Fern{\'a}ndez-Pend{\'a}s}, \citenamefont {Warren}, \citenamefont {Kosen}, \citenamefont {Scigliuzzo}, \citenamefont {Frisk~Kockum}, \citenamefont {Tancredi}, \citenamefont {Fadavi~Roudsari},\ and\ \citenamefont {Bylander}}]{osman2023mitigation}%
  \BibitemOpen
  \bibfield  {author} {\bibinfo {author} {\bibfnamefont {A.}~\bibnamefont {Osman}}, \bibinfo {author} {\bibfnamefont {J.}~\bibnamefont {Fern{\'a}ndez-Pend{\'a}s}}, \bibinfo {author} {\bibfnamefont {C.}~\bibnamefont {Warren}}, \bibinfo {author} {\bibfnamefont {S.}~\bibnamefont {Kosen}}, \bibinfo {author} {\bibfnamefont {M.}~\bibnamefont {Scigliuzzo}}, \bibinfo {author} {\bibfnamefont {A.}~\bibnamefont {Frisk~Kockum}}, \bibinfo {author} {\bibfnamefont {G.}~\bibnamefont {Tancredi}}, \bibinfo {author} {\bibfnamefont {A.}~\bibnamefont {Fadavi~Roudsari}},\ and\ \bibinfo {author} {\bibfnamefont {J.}~\bibnamefont {Bylander}},\ }\bibfield  {title} {\bibinfo {title} {Mitigation of frequency collisions in superconducting quantum processors},\ }\href@noop {} {\bibfield  {journal} {\bibinfo  {journal} {Physical Review Research}\ }\textbf {\bibinfo {volume} {5}},\ \bibinfo {pages} {043001} (\bibinfo {year} {2023})}\BibitemShut {NoStop}%
\bibitem [{\citenamefont {Acharya}\ \emph {et~al.}(2024)\citenamefont {Acharya}, \citenamefont {Armstrong}, \citenamefont {Balaji}, \citenamefont {Crawford}, \citenamefont {Gates}, \citenamefont {Gow}, \citenamefont {Kennedy}, \citenamefont {Pothuraju}, \citenamefont {Shahbazi},\ and\ \citenamefont {Shelly}}]{acharya2024integration}%
  \BibitemOpen
  \bibfield  {author} {\bibinfo {author} {\bibfnamefont {N.}~\bibnamefont {Acharya}}, \bibinfo {author} {\bibfnamefont {R.}~\bibnamefont {Armstrong}}, \bibinfo {author} {\bibfnamefont {Y.}~\bibnamefont {Balaji}}, \bibinfo {author} {\bibfnamefont {K.~G.}\ \bibnamefont {Crawford}}, \bibinfo {author} {\bibfnamefont {J.~C.}\ \bibnamefont {Gates}}, \bibinfo {author} {\bibfnamefont {P.~C.}\ \bibnamefont {Gow}}, \bibinfo {author} {\bibfnamefont {O.~W.}\ \bibnamefont {Kennedy}}, \bibinfo {author} {\bibfnamefont {R.~D.}\ \bibnamefont {Pothuraju}}, \bibinfo {author} {\bibfnamefont {K.}~\bibnamefont {Shahbazi}},\ and\ \bibinfo {author} {\bibfnamefont {C.~D.}\ \bibnamefont {Shelly}},\ }\bibfield  {title} {\bibinfo {title} {Integration of through-sapphire substrate machining with superconducting quantum processors},\ }\href@noop {} {\bibfield  {journal} {\bibinfo  {journal} {Advanced Materials}\ ,\ \bibinfo {pages} {2411780}} (\bibinfo {year} {2024})}\BibitemShut {NoStop}%
\bibitem [{\citenamefont {Van~Damme}\ \emph {et~al.}(2024)\citenamefont {Van~Damme}, \citenamefont {Massar}, \citenamefont {Acharya}, \citenamefont {Ivanov}, \citenamefont {Perez~Lozano}, \citenamefont {Canvel}, \citenamefont {Demarets}, \citenamefont {Vangoidsenhoven}, \citenamefont {Hermans}, \citenamefont {Lai} \emph {et~al.}}]{van2024advanced}%
  \BibitemOpen
  \bibfield  {author} {\bibinfo {author} {\bibfnamefont {J.}~\bibnamefont {Van~Damme}}, \bibinfo {author} {\bibfnamefont {S.}~\bibnamefont {Massar}}, \bibinfo {author} {\bibfnamefont {R.}~\bibnamefont {Acharya}}, \bibinfo {author} {\bibfnamefont {T.}~\bibnamefont {Ivanov}}, \bibinfo {author} {\bibfnamefont {D.}~\bibnamefont {Perez~Lozano}}, \bibinfo {author} {\bibfnamefont {Y.}~\bibnamefont {Canvel}}, \bibinfo {author} {\bibfnamefont {M.}~\bibnamefont {Demarets}}, \bibinfo {author} {\bibfnamefont {D.}~\bibnamefont {Vangoidsenhoven}}, \bibinfo {author} {\bibfnamefont {Y.}~\bibnamefont {Hermans}}, \bibinfo {author} {\bibfnamefont {J.}~\bibnamefont {Lai}}, \emph {et~al.},\ }\bibfield  {title} {\bibinfo {title} {Advanced cmos manufacturing of superconducting qubits on 300 mm wafers},\ }\href@noop {} {\bibfield  {journal} {\bibinfo  {journal} {Nature}\ ,\ \bibinfo {pages} {1}} (\bibinfo {year} {2024})}\BibitemShut {NoStop}%
\bibitem [{\citenamefont {Josephson}(1962)}]{josephson1962possible}%
  \BibitemOpen
  \bibfield  {author} {\bibinfo {author} {\bibfnamefont {B.~D.}\ \bibnamefont {Josephson}},\ }\bibfield  {title} {\bibinfo {title} {Possible new effects in superconductive tunnelling},\ }\href@noop {} {\bibfield  {journal} {\bibinfo  {journal} {Physics letters}\ }\textbf {\bibinfo {volume} {1}},\ \bibinfo {pages} {251} (\bibinfo {year} {1962})}\BibitemShut {NoStop}%
\bibitem [{\citenamefont {Golubov}\ \emph {et~al.}(2004)\citenamefont {Golubov}, \citenamefont {Kupriyanov},\ and\ \citenamefont {Il’Ichev}}]{golubov2004current}%
  \BibitemOpen
  \bibfield  {author} {\bibinfo {author} {\bibfnamefont {A.~A.}\ \bibnamefont {Golubov}}, \bibinfo {author} {\bibfnamefont {M.~Y.}\ \bibnamefont {Kupriyanov}},\ and\ \bibinfo {author} {\bibfnamefont {E.}~\bibnamefont {Il’Ichev}},\ }\bibfield  {title} {\bibinfo {title} {The current-phase relation in josephson junctions},\ }\href@noop {} {\bibfield  {journal} {\bibinfo  {journal} {Reviews of modern physics}\ }\textbf {\bibinfo {volume} {76}},\ \bibinfo {pages} {411} (\bibinfo {year} {2004})}\BibitemShut {NoStop}%
\bibitem [{\citenamefont {Bayros}\ \emph {et~al.}(2024)\citenamefont {Bayros}, \citenamefont {Cyster}, \citenamefont {Smith},\ and\ \citenamefont {Cole}}]{bayros2024influence}%
  \BibitemOpen
  \bibfield  {author} {\bibinfo {author} {\bibfnamefont {K.}~\bibnamefont {Bayros}}, \bibinfo {author} {\bibfnamefont {M.}~\bibnamefont {Cyster}}, \bibinfo {author} {\bibfnamefont {J.}~\bibnamefont {Smith}},\ and\ \bibinfo {author} {\bibfnamefont {J.}~\bibnamefont {Cole}},\ }\bibfield  {title} {\bibinfo {title} {Influence of pinholes and weak-points in aluminum-oxide josephson junctions},\ }\href@noop {} {\bibfield  {journal} {\bibinfo  {journal} {Physical Review Materials}\ }\textbf {\bibinfo {volume} {8}},\ \bibinfo {pages} {046202} (\bibinfo {year} {2024})}\BibitemShut {NoStop}%
\bibitem [{\citenamefont {Willsch}\ \emph {et~al.}(2024)\citenamefont {Willsch}, \citenamefont {Rieger}, \citenamefont {Winkel}, \citenamefont {Willsch}, \citenamefont {Dickel}, \citenamefont {Krause}, \citenamefont {Ando}, \citenamefont {Lescanne}, \citenamefont {Leghtas}, \citenamefont {Bronn} \emph {et~al.}}]{willsch2024observation}%
  \BibitemOpen
  \bibfield  {author} {\bibinfo {author} {\bibfnamefont {D.}~\bibnamefont {Willsch}}, \bibinfo {author} {\bibfnamefont {D.}~\bibnamefont {Rieger}}, \bibinfo {author} {\bibfnamefont {P.}~\bibnamefont {Winkel}}, \bibinfo {author} {\bibfnamefont {M.}~\bibnamefont {Willsch}}, \bibinfo {author} {\bibfnamefont {C.}~\bibnamefont {Dickel}}, \bibinfo {author} {\bibfnamefont {J.}~\bibnamefont {Krause}}, \bibinfo {author} {\bibfnamefont {Y.}~\bibnamefont {Ando}}, \bibinfo {author} {\bibfnamefont {R.}~\bibnamefont {Lescanne}}, \bibinfo {author} {\bibfnamefont {Z.}~\bibnamefont {Leghtas}}, \bibinfo {author} {\bibfnamefont {N.~T.}\ \bibnamefont {Bronn}}, \emph {et~al.},\ }\bibfield  {title} {\bibinfo {title} {Observation of josephson harmonics in tunnel junctions},\ }\href@noop {} {\bibfield  {journal} {\bibinfo  {journal} {Nature Physics}\ ,\ \bibinfo {pages} {1}} (\bibinfo {year} {2024})}\BibitemShut {NoStop}%
\bibitem [{\citenamefont {Lee}\ \emph {et~al.}(2006)\citenamefont {Lee}, \citenamefont {Oh}, \citenamefont {Tseng}, \citenamefont {Jammy},\ and\ \citenamefont {Huff}}]{lee2006gate}%
  \BibitemOpen
  \bibfield  {author} {\bibinfo {author} {\bibfnamefont {B.~H.}\ \bibnamefont {Lee}}, \bibinfo {author} {\bibfnamefont {J.}~\bibnamefont {Oh}}, \bibinfo {author} {\bibfnamefont {H.~H.}\ \bibnamefont {Tseng}}, \bibinfo {author} {\bibfnamefont {R.}~\bibnamefont {Jammy}},\ and\ \bibinfo {author} {\bibfnamefont {H.}~\bibnamefont {Huff}},\ }\bibfield  {title} {\bibinfo {title} {Gate stack technology for nanoscale devices},\ }\href@noop {} {\bibfield  {journal} {\bibinfo  {journal} {materials today}\ }\textbf {\bibinfo {volume} {9}},\ \bibinfo {pages} {32} (\bibinfo {year} {2006})}\BibitemShut {NoStop}%
\bibitem [{\citenamefont {Buchanan}\ and\ \citenamefont {Lo}(1997)}]{buchanan1997reliability}%
  \BibitemOpen
  \bibfield  {author} {\bibinfo {author} {\bibfnamefont {D.}~\bibnamefont {Buchanan}}\ and\ \bibinfo {author} {\bibfnamefont {S.-H.}\ \bibnamefont {Lo}},\ }\bibfield  {title} {\bibinfo {title} {Reliability and integration of ultra-thin gate dielectrics for advanced cmos},\ }\href@noop {} {\bibfield  {journal} {\bibinfo  {journal} {Microelectronic engineering}\ }\textbf {\bibinfo {volume} {36}},\ \bibinfo {pages} {13} (\bibinfo {year} {1997})}\BibitemShut {NoStop}%
\bibitem [{\citenamefont {Ekanayake}\ \emph {et~al.}(2004)\citenamefont {Ekanayake}, \citenamefont {Ford},\ and\ \citenamefont {Cortie}}]{ekanayake2004metal}%
  \BibitemOpen
  \bibfield  {author} {\bibinfo {author} {\bibfnamefont {S.}~\bibnamefont {Ekanayake}}, \bibinfo {author} {\bibfnamefont {M.}~\bibnamefont {Ford}},\ and\ \bibinfo {author} {\bibfnamefont {M.}~\bibnamefont {Cortie}},\ }\bibfield  {title} {\bibinfo {title} {Metal-insulator-metal (mim) nanocapacitors and effects of material properties on their operation},\ }\href@noop {} {\bibfield  {journal} {\bibinfo  {journal} {Materials Forum}\ }\textbf {\bibinfo {volume} {27}},\ \bibinfo {pages} {15} (\bibinfo {year} {2004})}\BibitemShut {NoStop}%
\bibitem [{\citenamefont {Da~Costa}\ \emph {et~al.}(2000)\citenamefont {Da~Costa}, \citenamefont {Tiusan}, \citenamefont {Dimopoulos},\ and\ \citenamefont {Ounadjela}}]{da2000tunneling}%
  \BibitemOpen
  \bibfield  {author} {\bibinfo {author} {\bibfnamefont {V.}~\bibnamefont {Da~Costa}}, \bibinfo {author} {\bibfnamefont {C.}~\bibnamefont {Tiusan}}, \bibinfo {author} {\bibfnamefont {T.}~\bibnamefont {Dimopoulos}},\ and\ \bibinfo {author} {\bibfnamefont {K.}~\bibnamefont {Ounadjela}},\ }\bibfield  {title} {\bibinfo {title} {Tunneling phenomena as a probe to investigate atomic scale fluctuations in metal/oxide/metal magnetic tunnel junctions},\ }\href@noop {} {\bibfield  {journal} {\bibinfo  {journal} {Physical Review Letters}\ }\textbf {\bibinfo {volume} {85}},\ \bibinfo {pages} {876} (\bibinfo {year} {2000})}\BibitemShut {NoStop}%
\bibitem [{\citenamefont {Zeng}\ \emph {et~al.}(2015)\citenamefont {Zeng}, \citenamefont {Nik}, \citenamefont {Greibe}, \citenamefont {Krantz}, \citenamefont {Wilson}, \citenamefont {Delsing},\ and\ \citenamefont {Olsson}}]{zeng2015direct}%
  \BibitemOpen
  \bibfield  {author} {\bibinfo {author} {\bibfnamefont {L.}~\bibnamefont {Zeng}}, \bibinfo {author} {\bibfnamefont {S.}~\bibnamefont {Nik}}, \bibinfo {author} {\bibfnamefont {T.}~\bibnamefont {Greibe}}, \bibinfo {author} {\bibfnamefont {P.}~\bibnamefont {Krantz}}, \bibinfo {author} {\bibfnamefont {C.}~\bibnamefont {Wilson}}, \bibinfo {author} {\bibfnamefont {P.}~\bibnamefont {Delsing}},\ and\ \bibinfo {author} {\bibfnamefont {E.}~\bibnamefont {Olsson}},\ }\bibfield  {title} {\bibinfo {title} {Direct observation of the thickness distribution of ultra thin alox barriers in al/alox/al josephson junctions},\ }\href@noop {} {\bibfield  {journal} {\bibinfo  {journal} {Journal of Physics D: Applied Physics}\ }\textbf {\bibinfo {volume} {48}},\ \bibinfo {pages} {395308} (\bibinfo {year} {2015})}\BibitemShut {NoStop}%
\bibitem [{\citenamefont {Dobrovinskaya}\ \emph {et~al.}(2009)\citenamefont {Dobrovinskaya}, \citenamefont {Lytvynov},\ and\ \citenamefont {Pishchik}}]{dobrovinskaya2009properties}%
  \BibitemOpen
  \bibfield  {author} {\bibinfo {author} {\bibfnamefont {E.~R.}\ \bibnamefont {Dobrovinskaya}}, \bibinfo {author} {\bibfnamefont {L.~A.}\ \bibnamefont {Lytvynov}},\ and\ \bibinfo {author} {\bibfnamefont {V.}~\bibnamefont {Pishchik}},\ }\bibfield  {title} {\bibinfo {title} {Properties of sapphire},\ }\href@noop {} {\bibfield  {journal} {\bibinfo  {journal} {Sapphire}\ ,\ \bibinfo {pages} {55}} (\bibinfo {year} {2009})}\BibitemShut {NoStop}%
\bibitem [{\citenamefont {Liu}\ \emph {et~al.}(2023)\citenamefont {Liu}, \citenamefont {Pan}, \citenamefont {Zhang},\ and\ \citenamefont {Feng}}]{liu2023unveiling}%
  \BibitemOpen
  \bibfield  {author} {\bibinfo {author} {\bibfnamefont {X.}~\bibnamefont {Liu}}, \bibinfo {author} {\bibfnamefont {K.}~\bibnamefont {Pan}}, \bibinfo {author} {\bibfnamefont {Z.}~\bibnamefont {Zhang}},\ and\ \bibinfo {author} {\bibfnamefont {Z.}~\bibnamefont {Feng}},\ }\bibfield  {title} {\bibinfo {title} {Unveiling atomic structure and chemical composition of the al/alox/al josephson junctions in qubits},\ }\href@noop {} {\bibfield  {journal} {\bibinfo  {journal} {Applied Surface Science}\ }\textbf {\bibinfo {volume} {640}},\ \bibinfo {pages} {158337} (\bibinfo {year} {2023})}\BibitemShut {NoStop}%
\bibitem [{\citenamefont {Supple}\ \emph {et~al.}(2021)\citenamefont {Supple}, \citenamefont {Holtz}, \citenamefont {Richardson},\ and\ \citenamefont {Gorman}}]{supple2021atomic}%
  \BibitemOpen
  \bibfield  {author} {\bibinfo {author} {\bibfnamefont {E.}~\bibnamefont {Supple}}, \bibinfo {author} {\bibfnamefont {M.}~\bibnamefont {Holtz}}, \bibinfo {author} {\bibfnamefont {C.~J.}\ \bibnamefont {Richardson}},\ and\ \bibinfo {author} {\bibfnamefont {B.}~\bibnamefont {Gorman}},\ }\bibfield  {title} {\bibinfo {title} {Atomic structure of superconducting tunnel junctions using stem and apt},\ }\href@noop {} {\bibfield  {journal} {\bibinfo  {journal} {Microscopy and Microanalysis}\ }\textbf {\bibinfo {volume} {27}},\ \bibinfo {pages} {2460} (\bibinfo {year} {2021})}\BibitemShut {NoStop}%
\bibitem [{\citenamefont {Cyster}\ \emph {et~al.}(2021)\citenamefont {Cyster}, \citenamefont {Smith}, \citenamefont {Vogt}, \citenamefont {Opletal}, \citenamefont {Russo},\ and\ \citenamefont {Cole}}]{cyster2021simulating}%
  \BibitemOpen
  \bibfield  {author} {\bibinfo {author} {\bibfnamefont {M.}~\bibnamefont {Cyster}}, \bibinfo {author} {\bibfnamefont {J.}~\bibnamefont {Smith}}, \bibinfo {author} {\bibfnamefont {N.}~\bibnamefont {Vogt}}, \bibinfo {author} {\bibfnamefont {G.}~\bibnamefont {Opletal}}, \bibinfo {author} {\bibfnamefont {S.}~\bibnamefont {Russo}},\ and\ \bibinfo {author} {\bibfnamefont {J.}~\bibnamefont {Cole}},\ }\bibfield  {title} {\bibinfo {title} {Simulating the fabrication of aluminium oxide tunnel junctions},\ }\href@noop {} {\bibfield  {journal} {\bibinfo  {journal} {npj quantum information}\ }\textbf {\bibinfo {volume} {7}},\ \bibinfo {pages} {12} (\bibinfo {year} {2021})}\BibitemShut {NoStop}%
\bibitem [{\citenamefont {Palumbo}\ \emph {et~al.}(2020)\citenamefont {Palumbo}, \citenamefont {Wen}, \citenamefont {Lombardo}, \citenamefont {Pazos}, \citenamefont {Aguirre}, \citenamefont {Eizenberg}, \citenamefont {Hui},\ and\ \citenamefont {Lanza}}]{palumbo2020review}%
  \BibitemOpen
  \bibfield  {author} {\bibinfo {author} {\bibfnamefont {F.}~\bibnamefont {Palumbo}}, \bibinfo {author} {\bibfnamefont {C.}~\bibnamefont {Wen}}, \bibinfo {author} {\bibfnamefont {S.}~\bibnamefont {Lombardo}}, \bibinfo {author} {\bibfnamefont {S.}~\bibnamefont {Pazos}}, \bibinfo {author} {\bibfnamefont {F.}~\bibnamefont {Aguirre}}, \bibinfo {author} {\bibfnamefont {M.}~\bibnamefont {Eizenberg}}, \bibinfo {author} {\bibfnamefont {F.}~\bibnamefont {Hui}},\ and\ \bibinfo {author} {\bibfnamefont {M.}~\bibnamefont {Lanza}},\ }\bibfield  {title} {\bibinfo {title} {A review on dielectric breakdown in thin dielectrics: silicon dioxide, high-k, and layered dielectrics},\ }\href@noop {} {\bibfield  {journal} {\bibinfo  {journal} {Advanced Functional Materials}\ }\textbf {\bibinfo {volume} {30}},\ \bibinfo {pages} {1900657} (\bibinfo {year} {2020})}\BibitemShut {NoStop}%
\bibitem [{\citenamefont {Dolan}(1977)}]{dolan1977offset}%
  \BibitemOpen
  \bibfield  {author} {\bibinfo {author} {\bibfnamefont {G.}~\bibnamefont {Dolan}},\ }\bibfield  {title} {\bibinfo {title} {Offset masks for lift-off photoprocessing},\ }\href@noop {} {\bibfield  {journal} {\bibinfo  {journal} {Applied Physics Letters}\ }\textbf {\bibinfo {volume} {31}},\ \bibinfo {pages} {337} (\bibinfo {year} {1977})}\BibitemShut {NoStop}%
\bibitem [{\citenamefont {Simmons}(1963)}]{simmons1963generalized}%
  \BibitemOpen
  \bibfield  {author} {\bibinfo {author} {\bibfnamefont {J.~G.}\ \bibnamefont {Simmons}},\ }\bibfield  {title} {\bibinfo {title} {Generalized formula for the electric tunnel effect between similar electrodes separated by a thin insulating film},\ }\href@noop {} {\bibfield  {journal} {\bibinfo  {journal} {Journal of applied physics}\ }\textbf {\bibinfo {volume} {34}},\ \bibinfo {pages} {1793} (\bibinfo {year} {1963})}\BibitemShut {NoStop}%
\bibitem [{\citenamefont {Koberidze}\ \emph {et~al.}(2016)\citenamefont {Koberidze}, \citenamefont {Feshchenko}, \citenamefont {Puska}, \citenamefont {Nieminen},\ and\ \citenamefont {Pekola}}]{koberidze2016effect}%
  \BibitemOpen
  \bibfield  {author} {\bibinfo {author} {\bibfnamefont {M.}~\bibnamefont {Koberidze}}, \bibinfo {author} {\bibfnamefont {A.}~\bibnamefont {Feshchenko}}, \bibinfo {author} {\bibfnamefont {M.}~\bibnamefont {Puska}}, \bibinfo {author} {\bibfnamefont {R.}~\bibnamefont {Nieminen}},\ and\ \bibinfo {author} {\bibfnamefont {J.}~\bibnamefont {Pekola}},\ }\bibfield  {title} {\bibinfo {title} {Effect of interface geometry on electron tunnelling in al/al2o3/al junctions},\ }\href@noop {} {\bibfield  {journal} {\bibinfo  {journal} {Journal of Physics D: Applied Physics}\ }\textbf {\bibinfo {volume} {49}},\ \bibinfo {pages} {165303} (\bibinfo {year} {2016})}\BibitemShut {NoStop}%
\bibitem [{\citenamefont {Koberidze}\ \emph {et~al.}(2018)\citenamefont {Koberidze}, \citenamefont {Puska},\ and\ \citenamefont {Nieminen}}]{koberidze2018structural}%
  \BibitemOpen
  \bibfield  {author} {\bibinfo {author} {\bibfnamefont {M.}~\bibnamefont {Koberidze}}, \bibinfo {author} {\bibfnamefont {M.}~\bibnamefont {Puska}},\ and\ \bibinfo {author} {\bibfnamefont {R.}~\bibnamefont {Nieminen}},\ }\bibfield  {title} {\bibinfo {title} {Structural details of al/al 2 o 3 junctions and their role in the formation of electron tunnel barriers},\ }\href@noop {} {\bibfield  {journal} {\bibinfo  {journal} {Physical Review B}\ }\textbf {\bibinfo {volume} {97}},\ \bibinfo {pages} {195406} (\bibinfo {year} {2018})}\BibitemShut {NoStop}%
\bibitem [{\citenamefont {Dorneles}\ \emph {et~al.}(2003)\citenamefont {Dorneles}, \citenamefont {Schaefer}, \citenamefont {Carara},\ and\ \citenamefont {Schelp}}]{dorneles2003use}%
  \BibitemOpen
  \bibfield  {author} {\bibinfo {author} {\bibfnamefont {L.}~\bibnamefont {Dorneles}}, \bibinfo {author} {\bibfnamefont {D.}~\bibnamefont {Schaefer}}, \bibinfo {author} {\bibfnamefont {M.}~\bibnamefont {Carara}},\ and\ \bibinfo {author} {\bibfnamefont {L.}~\bibnamefont {Schelp}},\ }\bibfield  {title} {\bibinfo {title} {The use of simmons’ equation to quantify the insulating barrier parameters in al/alo x/al tunnel junctions},\ }\href@noop {} {\bibfield  {journal} {\bibinfo  {journal} {Applied physics letters}\ }\textbf {\bibinfo {volume} {82}},\ \bibinfo {pages} {2832} (\bibinfo {year} {2003})}\BibitemShut {NoStop}%
\bibitem [{\citenamefont {Kim}\ \emph {et~al.}(2020)\citenamefont {Kim}, \citenamefont {Ray},\ and\ \citenamefont {Lordi}}]{kim2020density}%
  \BibitemOpen
  \bibfield  {author} {\bibinfo {author} {\bibfnamefont {C.-E.}\ \bibnamefont {Kim}}, \bibinfo {author} {\bibfnamefont {K.~G.}\ \bibnamefont {Ray}},\ and\ \bibinfo {author} {\bibfnamefont {V.}~\bibnamefont {Lordi}},\ }\bibfield  {title} {\bibinfo {title} {A density-functional theory study of the al/alox/al tunnel junction},\ }\href@noop {} {\bibfield  {journal} {\bibinfo  {journal} {Journal of applied physics}\ }\textbf {\bibinfo {volume} {128}} (\bibinfo {year} {2020})}\BibitemShut {NoStop}%
\bibitem [{\citenamefont {Snijders}\ \emph {et~al.}(2002)\citenamefont {Snijders}, \citenamefont {Jeurgens},\ and\ \citenamefont {Sloof}}]{snijders2002structure}%
  \BibitemOpen
  \bibfield  {author} {\bibinfo {author} {\bibfnamefont {P.}~\bibnamefont {Snijders}}, \bibinfo {author} {\bibfnamefont {L.}~\bibnamefont {Jeurgens}},\ and\ \bibinfo {author} {\bibfnamefont {W.}~\bibnamefont {Sloof}},\ }\bibfield  {title} {\bibinfo {title} {Structure of thin aluminium-oxide films determined from valence band spectra measured using xps},\ }\href@noop {} {\bibfield  {journal} {\bibinfo  {journal} {Surface science}\ }\textbf {\bibinfo {volume} {496}},\ \bibinfo {pages} {97} (\bibinfo {year} {2002})}\BibitemShut {NoStop}%
\bibitem [{\citenamefont {Fritz}\ \emph {et~al.}(2019{\natexlab{b}})\citenamefont {Fritz}, \citenamefont {Radtke}, \citenamefont {Schneider}, \citenamefont {Weides},\ and\ \citenamefont {Gerthsen}}]{fritz2019optimization}%
  \BibitemOpen
  \bibfield  {author} {\bibinfo {author} {\bibfnamefont {S.}~\bibnamefont {Fritz}}, \bibinfo {author} {\bibfnamefont {L.}~\bibnamefont {Radtke}}, \bibinfo {author} {\bibfnamefont {R.}~\bibnamefont {Schneider}}, \bibinfo {author} {\bibfnamefont {M.}~\bibnamefont {Weides}},\ and\ \bibinfo {author} {\bibfnamefont {D.}~\bibnamefont {Gerthsen}},\ }\bibfield  {title} {\bibinfo {title} {Optimization of al/alox/al-layer systems for josephson junctions from a microstructure point of view},\ }\href@noop {} {\bibfield  {journal} {\bibinfo  {journal} {Journal of Applied Physics}\ }\textbf {\bibinfo {volume} {125}} (\bibinfo {year} {2019}{\natexlab{b}})}\BibitemShut {NoStop}%
\bibitem [{\citenamefont {Oh}\ \emph {et~al.}(2025)\citenamefont {Oh}, \citenamefont {Kopas}, \citenamefont {Cansizoglu}, \citenamefont {Mutus}, \citenamefont {Yadavalli}, \citenamefont {Kim}, \citenamefont {Kramer}, \citenamefont {King},\ and\ \citenamefont {Zhou}}]{oh2025correlating}%
  \BibitemOpen
  \bibfield  {author} {\bibinfo {author} {\bibfnamefont {J.-S.}\ \bibnamefont {Oh}}, \bibinfo {author} {\bibfnamefont {C.~J.}\ \bibnamefont {Kopas}}, \bibinfo {author} {\bibfnamefont {H.}~\bibnamefont {Cansizoglu}}, \bibinfo {author} {\bibfnamefont {J.~Y.}\ \bibnamefont {Mutus}}, \bibinfo {author} {\bibfnamefont {K.}~\bibnamefont {Yadavalli}}, \bibinfo {author} {\bibfnamefont {T.-H.}\ \bibnamefont {Kim}}, \bibinfo {author} {\bibfnamefont {M.}~\bibnamefont {Kramer}}, \bibinfo {author} {\bibfnamefont {A.~H.}\ \bibnamefont {King}},\ and\ \bibinfo {author} {\bibfnamefont {L.}~\bibnamefont {Zhou}},\ }\bibfield  {title} {\bibinfo {title} {Correlating aluminum layer deposition rates, josephson junction microstructure, and superconducting qubits’ performance},\ }\href@noop {} {\bibfield  {journal} {\bibinfo  {journal} {Acta Materialia}\ }\textbf {\bibinfo {volume} {284}},\ \bibinfo {pages} {120631} (\bibinfo {year} {2025})}\BibitemShut {NoStop}%
\bibitem [{\citenamefont {Strand}\ and\ \citenamefont {Shluger}(2024)}]{strand2024structure}%
  \BibitemOpen
  \bibfield  {author} {\bibinfo {author} {\bibfnamefont {J.}~\bibnamefont {Strand}}\ and\ \bibinfo {author} {\bibfnamefont {A.~L.}\ \bibnamefont {Shluger}},\ }\bibfield  {title} {\bibinfo {title} {On the structure of oxygen deficient amorphous oxide films},\ }\href@noop {} {\bibfield  {journal} {\bibinfo  {journal} {Advanced Science}\ }\textbf {\bibinfo {volume} {11}},\ \bibinfo {pages} {2306243} (\bibinfo {year} {2024})}\BibitemShut {NoStop}%
\bibitem [{Note1()}]{Note1}%
  \BibitemOpen
  \bibinfo {note} {0.2~nm approximates the ionic radii in the AlO$_{\protect \rm x}$ barrier}\BibitemShut {NoStop}%
\bibitem [{\citenamefont {Kolodzey}\ \emph {et~al.}(2000)\citenamefont {Kolodzey}, \citenamefont {Chowdhury}, \citenamefont {Adam}, \citenamefont {Qui}, \citenamefont {Rau}, \citenamefont {Olowolafe}, \citenamefont {Suehle},\ and\ \citenamefont {Chen}}]{kolodzey2000electrical}%
  \BibitemOpen
  \bibfield  {author} {\bibinfo {author} {\bibfnamefont {J.}~\bibnamefont {Kolodzey}}, \bibinfo {author} {\bibfnamefont {E.~A.}\ \bibnamefont {Chowdhury}}, \bibinfo {author} {\bibfnamefont {T.~N.}\ \bibnamefont {Adam}}, \bibinfo {author} {\bibfnamefont {G.}~\bibnamefont {Qui}}, \bibinfo {author} {\bibfnamefont {I.}~\bibnamefont {Rau}}, \bibinfo {author} {\bibfnamefont {J.~O.}\ \bibnamefont {Olowolafe}}, \bibinfo {author} {\bibfnamefont {J.~S.}\ \bibnamefont {Suehle}},\ and\ \bibinfo {author} {\bibfnamefont {Y.}~\bibnamefont {Chen}},\ }\bibfield  {title} {\bibinfo {title} {Electrical conduction and dielectric breakdown in aluminum oxide insulators on silicon},\ }\href@noop {} {\bibfield  {journal} {\bibinfo  {journal} {IEEE Transactions on Electron Devices}\ }\textbf {\bibinfo {volume} {47}},\ \bibinfo {pages} {121} (\bibinfo {year} {2000})}\BibitemShut {NoStop}%
\bibitem [{\citenamefont {Aref}\ \emph {et~al.}(2014)\citenamefont {Aref}, \citenamefont {Averin}, \citenamefont {van Dijken}, \citenamefont {Ferring}, \citenamefont {Koberidze}, \citenamefont {Maisi}, \citenamefont {Nguyend}, \citenamefont {Nieminen}, \citenamefont {Pekola},\ and\ \citenamefont {Yao}}]{aref2014characterization}%
  \BibitemOpen
  \bibfield  {author} {\bibinfo {author} {\bibfnamefont {T.}~\bibnamefont {Aref}}, \bibinfo {author} {\bibfnamefont {A.}~\bibnamefont {Averin}}, \bibinfo {author} {\bibfnamefont {S.}~\bibnamefont {van Dijken}}, \bibinfo {author} {\bibfnamefont {A.}~\bibnamefont {Ferring}}, \bibinfo {author} {\bibfnamefont {M.}~\bibnamefont {Koberidze}}, \bibinfo {author} {\bibfnamefont {V.~F.}\ \bibnamefont {Maisi}}, \bibinfo {author} {\bibfnamefont {H.}~\bibnamefont {Nguyend}}, \bibinfo {author} {\bibfnamefont {R.~M.}\ \bibnamefont {Nieminen}}, \bibinfo {author} {\bibfnamefont {J.~P.}\ \bibnamefont {Pekola}},\ and\ \bibinfo {author} {\bibfnamefont {L.}~\bibnamefont {Yao}},\ }\bibfield  {title} {\bibinfo {title} {Characterization of aluminum oxide tunnel barriers by combining transport measurements and transmission electron microscopy imaging},\ }\href@noop {} {\bibfield  {journal} {\bibinfo  {journal} {Journal of Applied Physics}\ }\textbf {\bibinfo {volume} {116}} (\bibinfo {year} {2014})}\BibitemShut {NoStop}%
\bibitem [{\citenamefont {Bradski}(2000)}]{opencv_library}%
  \BibitemOpen
  \bibfield  {author} {\bibinfo {author} {\bibfnamefont {G.}~\bibnamefont {Bradski}},\ }\bibfield  {title} {\bibinfo {title} {{The OpenCV Library}},\ }\href@noop {} {\bibfield  {journal} {\bibinfo  {journal} {Dr. Dobb's Journal of Software Tools}\ } (\bibinfo {year} {2000})}\BibitemShut {NoStop}%
\bibitem [{Note2()}]{Note2}%
  \BibitemOpen
  \bibinfo {note} {Noise at the $\sim $0.6~nm range is present in some regions of these AFM scans, which we found hard to exclude from measurements. This represents $\sim $10~\% of the total topographic variation in these AFM maps and is not present across the whole scans and therefore we believe not a dominant contributor to the following analysis.}\BibitemShut {Stop}%
\bibitem [{\citenamefont {McGreevy}(2001)}]{mcgreevy2001reverse}%
  \BibitemOpen
  \bibfield  {author} {\bibinfo {author} {\bibfnamefont {R.~L.}\ \bibnamefont {McGreevy}},\ }\bibfield  {title} {\bibinfo {title} {Reverse monte carlo modelling},\ }\href@noop {} {\bibfield  {journal} {\bibinfo  {journal} {Journal of Physics: Condensed Matter}\ }\textbf {\bibinfo {volume} {13}},\ \bibinfo {pages} {R877} (\bibinfo {year} {2001})}\BibitemShut {NoStop}%
\end{thebibliography}%

\newpage

\appendix

\end{document}


\preprint{APS/123-QED}

\title{Supplementary Materials for Analysis of Josephson Junction Barrier Variation - a Combined Electron Microscopy, Breakdown and Monte-Carlo Approach}

\author{Oscar~W.~Kennedy}\email{okennedy@oqc.tech}
\author{Kevin~G.~Crawford}
\author{Kowsar~Shahbazi}
\author{Connor~D.~Shelly}\email{cshelly@oqc.tech}

\affiliation{%
 Oxford Quantum Circuits, Thames Valley Science Park, Shinfield, Reading, United Kingdom, RG2 9LH}%


\maketitle

\section{IV Fitting}
In the main text we discuss how we fit the Simmons model to the IVs at voltages lower than the breakdown voltage. The Simmons model has three input parameters, the area of the junction, the potential barrier height and the barrier thickness. In the main text we fix the area to that from AFM measurements and use the potential barrier height and barrier thickness as fit parameters. 

\begin{figure}
    \centering
    \includegraphics[width=\linewidth]{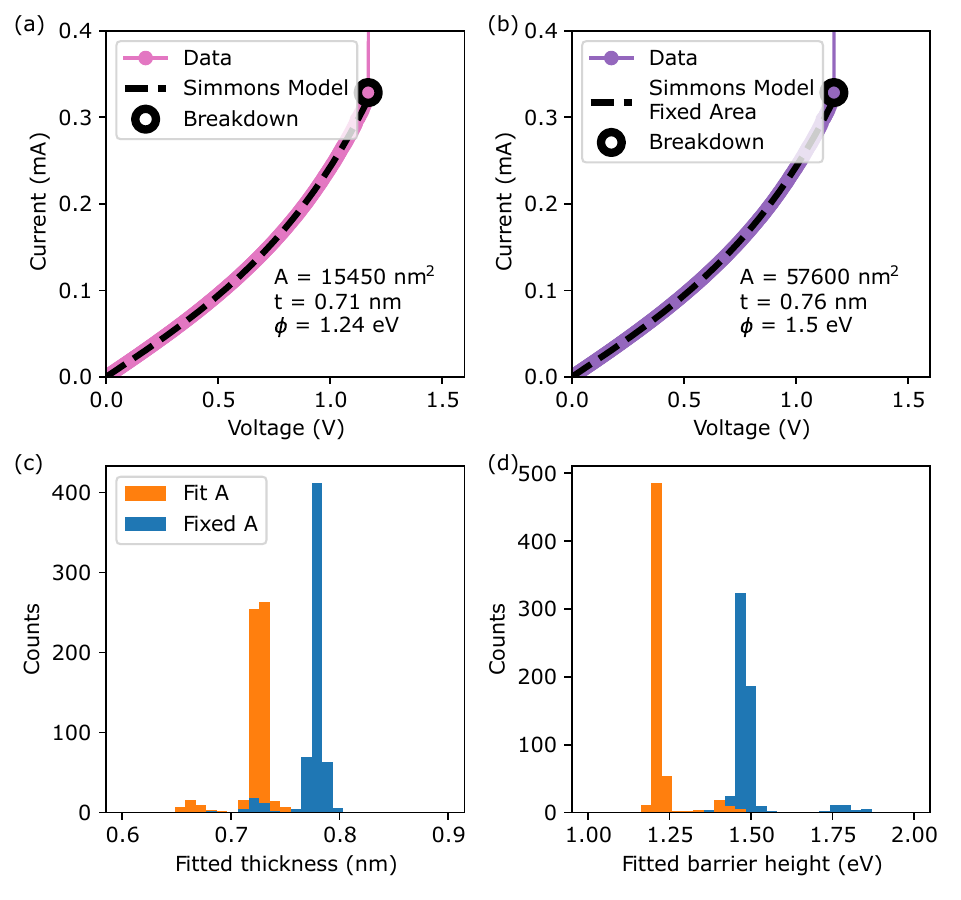}
    \caption{(a) Simmons model fit to an IV where A, t and $\phi$ are all fit parameters. (b) Simmons model fit to an IV where A is fixed to the value found by AFM and t and $\phi$ are fit parameters. (c) Histograms of fitted thickness comparing when A is a fit parameter or fixed by AFM. (d) Histograms of fitted barrier height comparing when A is a fit parameter or fixed by AFM.  }
    \label{fig:area_fix}
\end{figure}

In Fig.~\ref{fig:area_fix}~(a,b) we show IVs which have been fit to the Simmons model where the area is a fit parameter (a) and is fixed by AFM (b). Both curves are fit well by the Simmons model. We therefore choose to fix the area to AFM measurements in order to reduce our fit parameters and better constrain the model. In In Fig.~\ref{fig:area_fix}~(c,d) we show the effects of this on the distribution of fit parameters. Fixing the area increases both the average thickness and barrier height, but maintains the same shape of a dominant peak with a satellite peak. 

\section{AFM Measurements}
\begin{figure}[h]
    \centering
    \includegraphics[width=\linewidth]{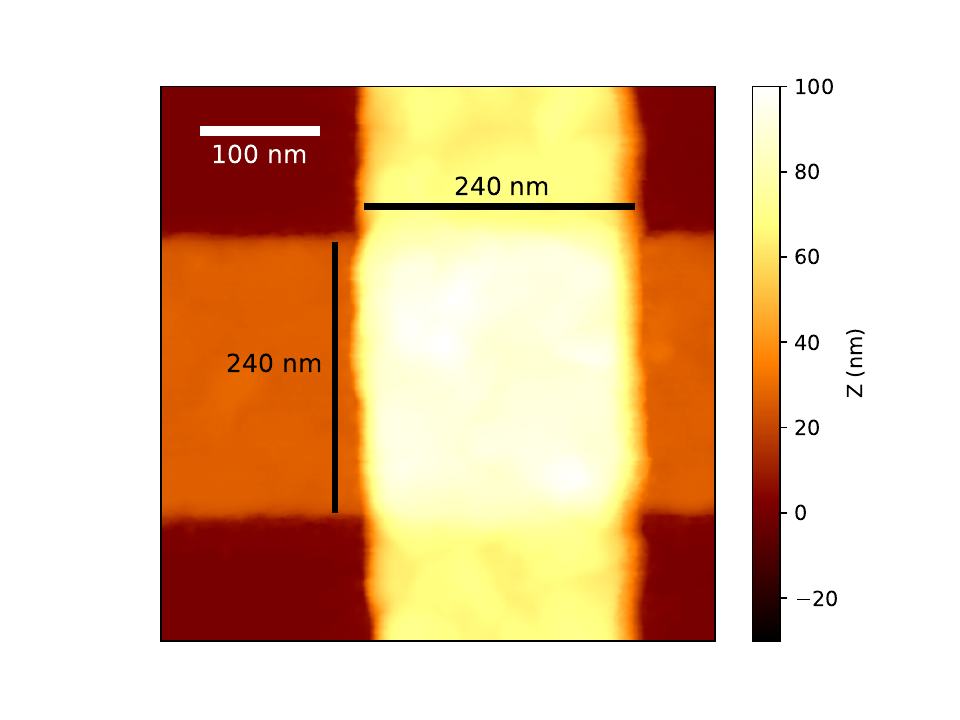}
    \caption{AFM image showing the 240 $\times$ 240~nm dimensions of the junction. }
    \label{fig:AFM-dimensions}
\end{figure}

In Fig.~\ref{fig:AFM-dimensions} we show an AFM map of a representative JJ. We show the $240 \times 240$~nm measurements used in the Monte Carlo simulations. 

\section{Additional STEM imaging}
\begin{figure}
    \centering
    \includegraphics[width=\linewidth]{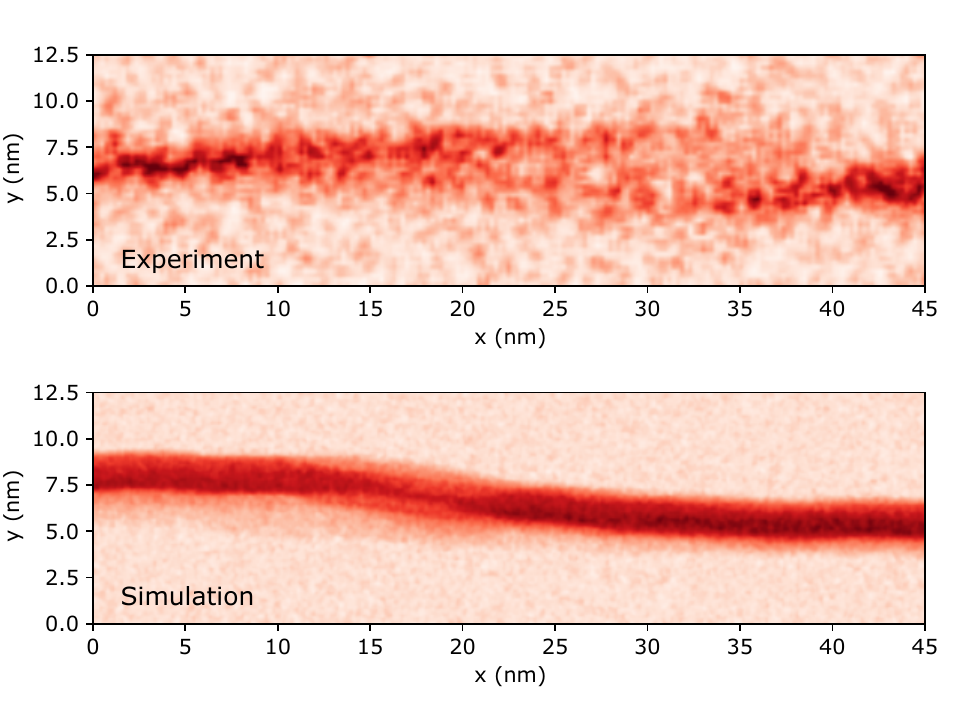}
    \caption{Comparison of (a) experimental and (b) simulated STEM EDS cross sections monitoring the oxygen peak across a barrier. The simulated STEM has been taken from the edge of a protrusion in the underlying bottom electrode.}
    \label{fig:stem_grain}
\end{figure}

In Fig.~\ref{fig:stem_grain} we compare (a) experimental and (b) simulated STEM profiles where the simulated STEM is taken from the edge of a protrusion in the bottom electrode. We see qualitatively similar features where the barrier appears to broaden whilst dropping in oxygen counts. The simulated barrier has been constructed to have a uniform thickness across and we understand that the broadening occurs due to the projection of the 3D barrier geometry into two dimensions. The operator could avoid regions that look like the type of feature shown here to try to minimise the amount of broadening that arises due to roughness in the bottom lead. 

We also provide two additional STEM images in Fig.~\ref{fig:extra_stem}, where the same kernel edge-detection protocol is used as in the main text resizing the kernel according to pixel size. These regions are larger the STEM image shown in the main text. These STEM images both have regions where there are features we identify with bottom-electrode roughness. The median thickness measurements for the higher resolution image are similar to those as found from the STEM image in the main text, although the standard deviation of the thicknesses is substantially increased because of the bottom electrodes. The lower resolution image shows a slightly increased average thickness, likely due to a combination of the reduced resolution and the substantial number of topographic features included in the imaged region. 

\begin{figure*}
    \includegraphics[width=0.65\linewidth]{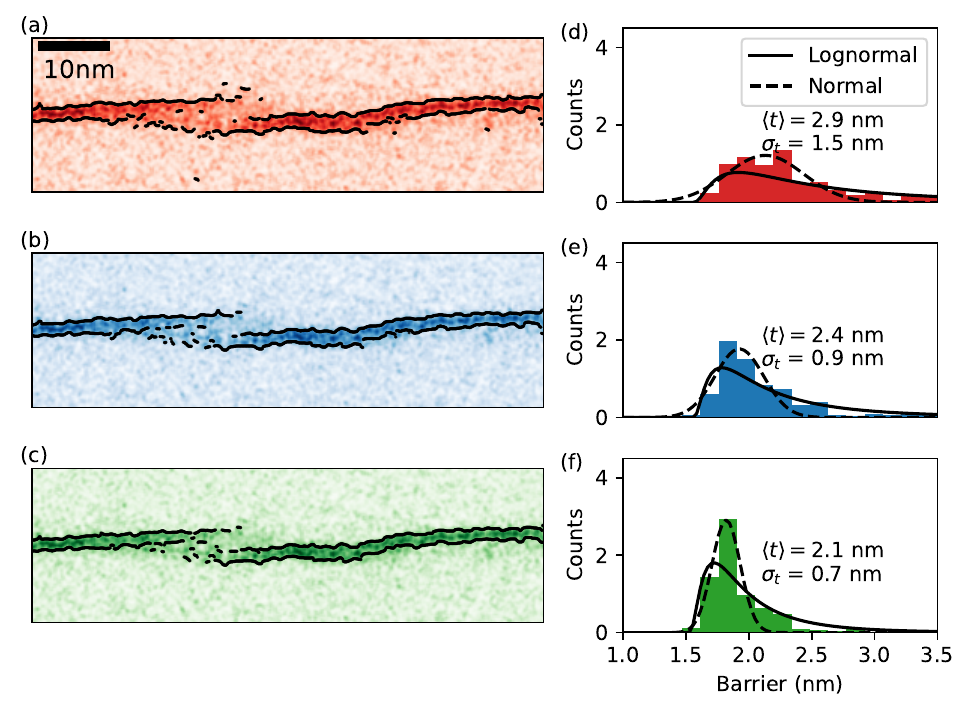}
    \caption{The first set of additional STEM images and kernel integration analysis complementing the figures in the main text. These areas include some regions where topographic features increase the inferred thickness distributions. (a-c) show heatmaps indicating the oxygen peak with the edges of the barrier as detected by Kernel integration overlaid to the maps. Different values of $\delta$ are used to define the kernel for the three maps (0, 0.2, 0.4 respectively). (d-f) show histograms of thicknesses as inferred from the different kernel integrations.}
    \label{fig:extra_stem}
\end{figure*}

\begin{figure*}
    \includegraphics[width=0.95\linewidth]{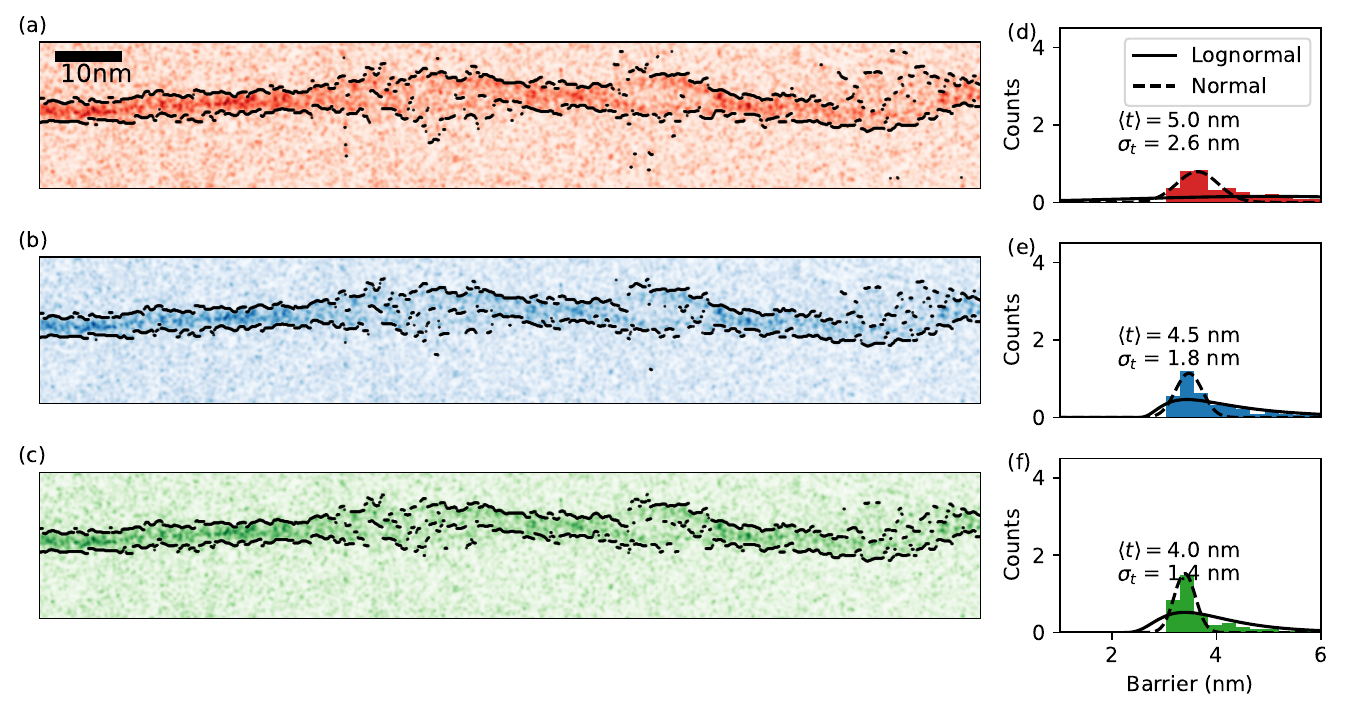}
    \caption{The second set of additional STEM images and kernel integration analysis complementing the figures in the main text. These areas include some regions where topographic features increase the inferred thickness distributions. (a-c) show heatmaps indicating the oxygen peak with the edges of the barrier as detected by Kernel integration overlaid to the maps. Different values of $\delta$ are used to define the kernel for the three maps (0, 0.2, 0.4 respectively). (d-f) show histograms of thicknesses as inferred from the different kernel integrations.}
    \label{fig:extra_stem}
\end{figure*}

\section{Spatial Variation}
\begin{figure*}
    \includegraphics[width=\linewidth]{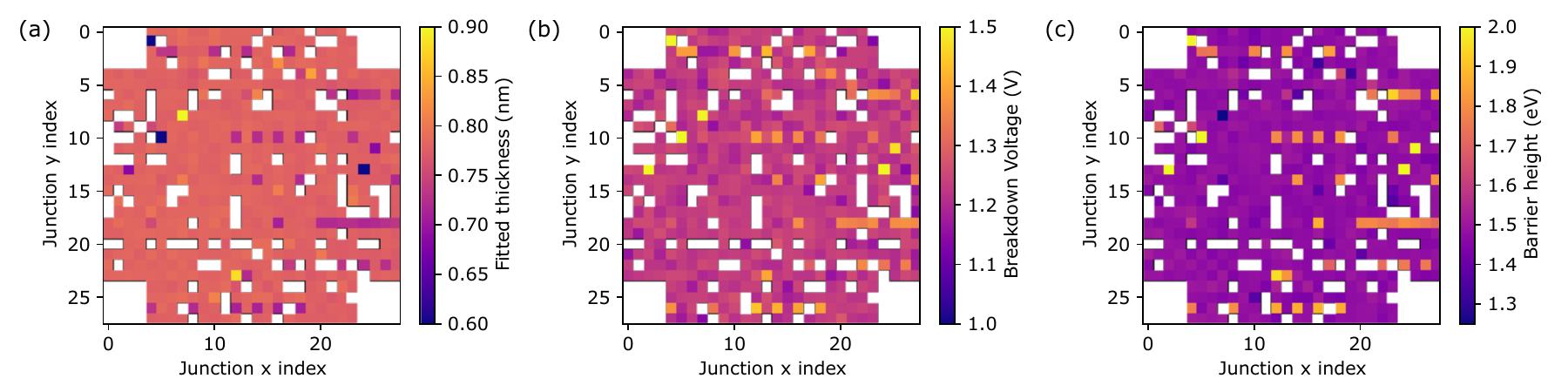}
    \caption{Semi-spatial plots of parameters from electrical characterisation. (a) Fitted thickness (nm) (b) Breakdown voltage (V) and (c) Barrier height (eV) as presented in the main text in histograms.}
    \label{fig:semi_spatial}
\end{figure*}
In order to investigate the bimodal distribution we see in breakdown voltage, we plot the spatial variation of parameters determined from electrical characterisation of junctions in Fig.~\ref{fig:semi_spatial}. The junctions are arrayed in tightly packed 4x4 grids which are separated by $\sim$~8~mm in x and y. We collapse this into a square grid to aid the visualisation of this data. 

We find that there are no large regions of the wafer which show aberrations in electrical values, although there are some clusters of points which do show aberrations close to one another (i.e. horizontal stripes). This may indicate that the second peak in the breakdown voltage arises due to some form of local defect. The key point remains, that the second sub-ensemble of junctions with different breakdown voltage are not something identified in a simple resistance measurement, but are obvious in a breakdown/Simmons fitting routine. This offers an opportunity to use this enhanced characterization as part of a fabrication optimisation process.

\section{Kernel Integration}

\begin{figure*}[h!]
    \centering
    \includegraphics[width=\linewidth]{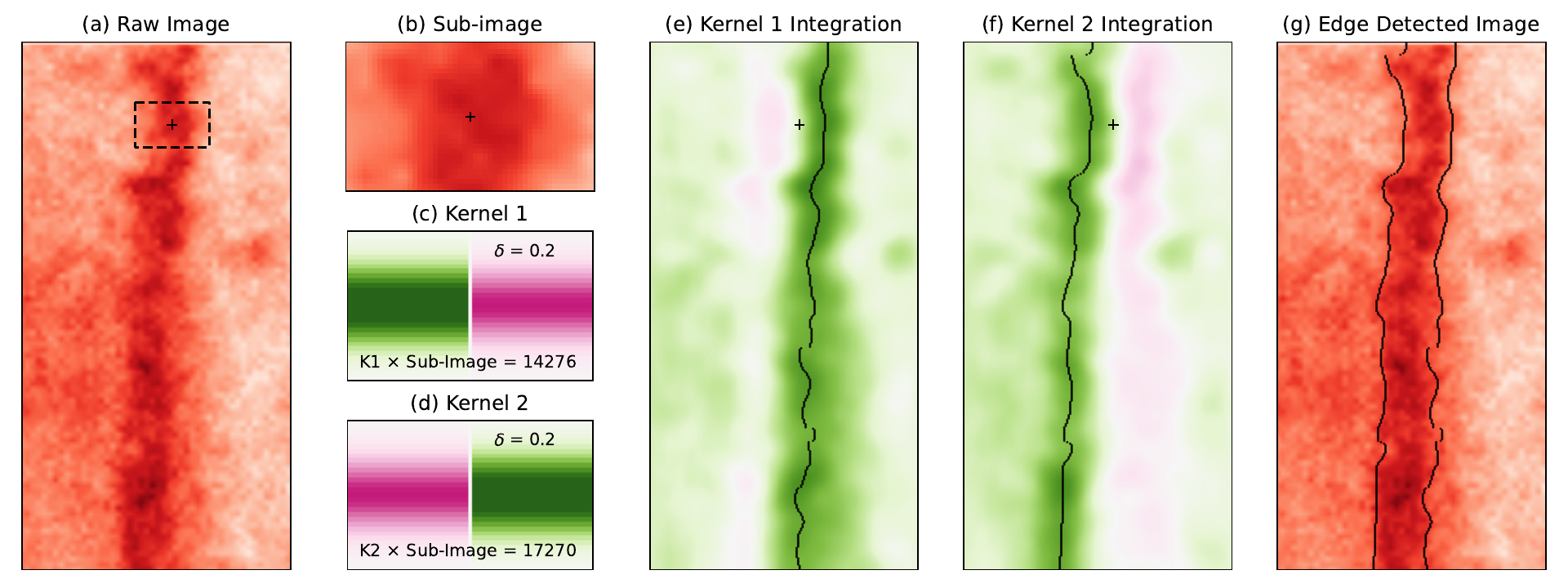}
    \caption{The kernel integration procedure is illustrated in this figure. We start with (a) a raw image showing, in this instance showing the STEM-EDS counts for a peak associated with oxygen. In the kernel integration a sub-image (b) is multiplied by a kernel (c, d) to give a value which represents a single value with the values of these pixels found by multiplying kernel 1 (c) and kernel 2 (d) by the sub-image in (b) are given in (c) and (d) respectively. (e, f) Show images resulting from rastering the kernels across the raw image, where the value of each pixel is the value from the multiplication of the kernel by a sub-image centred on the respective pixel from the raw image. On each row of pixels we can find the maximum value from the kernel integration and identify this as the edge. These are shown by the row of black markers in (e, f). (g) Shows the raw image with the two edges found by kernel integration overlaid.}
    \label{fig:Kernel}
\end{figure*}

In Fig.~4 of the main text we use kernel integration to identify the edges of an oxygen barrier. This technique involves splitting the original images into smaller sub-images and multiplying them by a `kernel' of the same dimensions to get a single value. With a suitable kernel, the single value can be dependent on a gradient in the values and so is suitable for our use case, where we want to identify the edges of an oxygen-rich region (i.e. regions of large gradients in oxygen content). 

To construct the kernels we first build a 1D array which is aligned along the thin axis of the barrier. At the centre of the 1D array we define a value of 0. We then build this out by appending $k$ instances of ($\delta + 1$) before the 0 and $k$ instances of ($\delta - 1$) after the 0, where $\delta$ is an asymmetry factor. The second kernel is defined by reflecting the first kernel about its centre (i.e. the pixel with value 0). 
We then add the second dimension of the kernel by adding weighted versions of the array described above to either side of the first array. We weight the array using a Gaussian function which decays with a length-scale of 0.5~nm. 
In Fig.~\ref{fig:Kernel} we show a break-down of how the kernel integration used in the main text is performed.

We demonstrate how the raw image containing information about the oxygen content, is divided into sub-images, which are in turn multiplied by a kernel to yield a single pixel value. We do this at each pixel in the raw image, generating a 2D image of kernel integration. The maximal values for each row in the kernel integration are identified as the edges of the barrier. 
The different edges shown in the main text are identified by changing the kernels used. We can introduce more or less asymmetry in the kernel by changing the value $\delta$ where this is described in the main text. 

\section{Additional Monte-Carlo Simulations}
\begin{figure*}[h!]
    \centering
    \includegraphics[width=\linewidth]{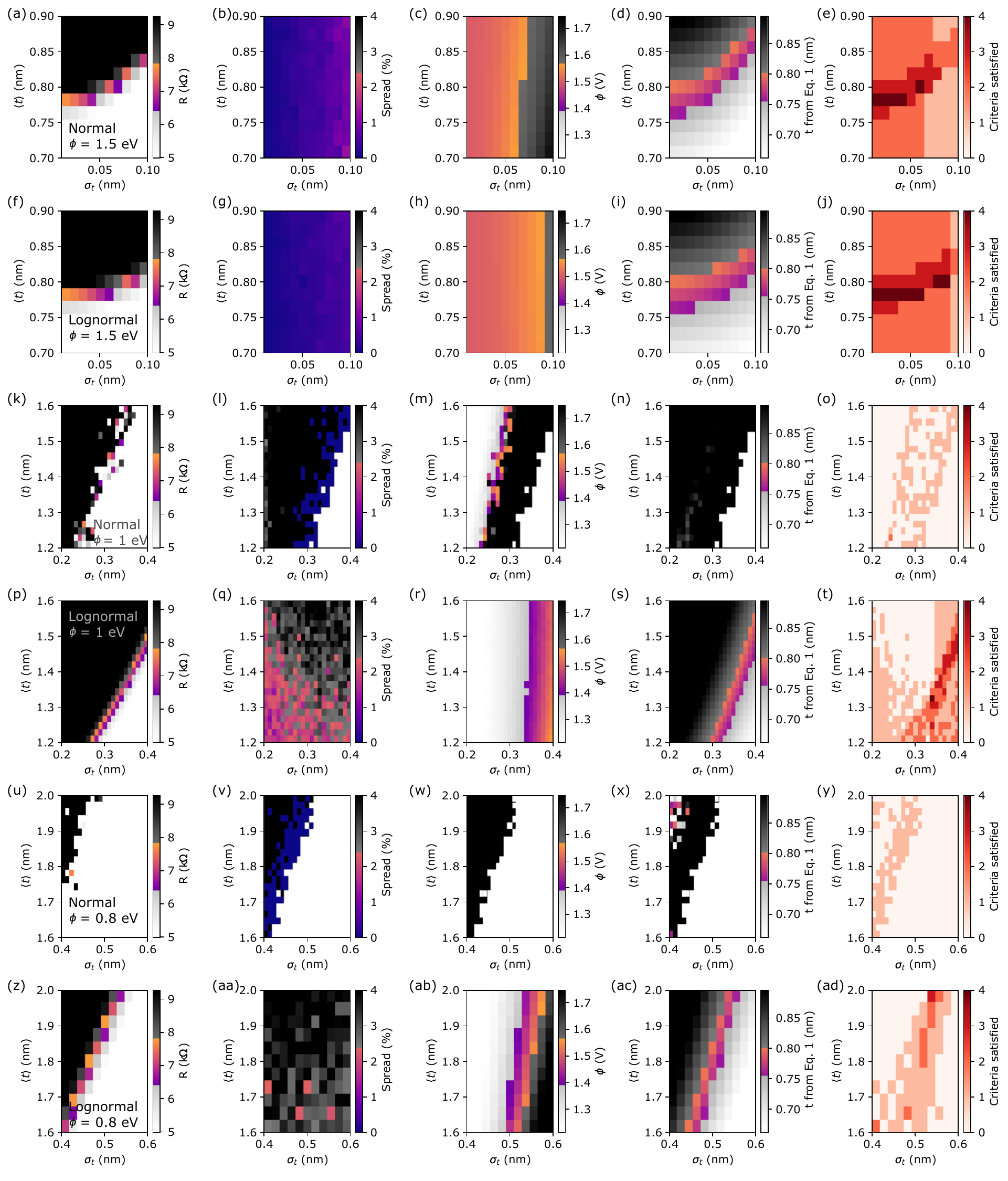}
    \caption{Results of Monte-Carlo simulations of JJs as in the main text. Results with a normally distributed barrier thickness are shown in (a-e), (k-o) and (u-y) and log-normally distributed thickness distribution in (f-j), (p-t) and (z-ad). 
    (a-j) Use a barrier height of 1.5~eV (k-n) use a barrier height of 1.0~eV and (u-ad) use a barrier height of 0.8~eV. There are regions where all criteria are satisfied for 1.5~eV, but the thickness variation is exceptionally small (less than the ionic radius of the constituent barrier material). For other distributions there are no regions where all parameters are satisfied. This supports the choice of the parameter regime of $\phi~\sim$~1.22~eV.}
    \label{fig:MC_IVs_SI}
\end{figure*}
In Fig.~\ref{fig:MC_IVs_SI} we show further Monte-Carlo simulations for barriers with lower barrier heights than shown in the main text (1.22~eV). 
In the top two rows we show results for normal and log-normal thickness distributions for a barrier height of 1~eV. In the normally distributed barrier we find a regime in the bottom right hand side of each panel with a thin barrier and a large standard deviation in the barrier thickness, where each junction simulated includes a short circuit (i.e. a thickness $\leq$ 0~nm) and is shown by a lack of colouring. 
The boundary to this region is characterised by a low resistance spread. This low resistance spread is caused by most junctions failing by short circuit and being excluded from calculations of spread. 
The boundary between resistances above the target resistance, and below the target resistance is not smooth. This is because a small number of pixels in the tail of the thickness distribution can contribute sufficiently to drops in resistance that this is not stable. It's possible that this would be avoided by using a finer pixel size. 

The log-normally distributed thickness distribution has a more stable map, where there is a clear boundary between resistance regimes. These are shown in the second and third rows in Fig.~\ref{fig:MC_IVs_SI} where we compute the results for 1~eV and 0.8~eV barrier heights respectively. As is obvious from Eq.1~main text, to maintain the same resistance, with a lower barrier height requires a thicker barrier. This is seen in the Monte-Carlo simulation. Using a barrier height of 0.8~eV, we get good matches between average fits of experimental IVs to the Simmons model and Monte-Carlo simulations with an average barrier height of $\sim$~2.1~nm and standard deviation in barrier thickness of $\sim$~0.55~nm. 
These average thicknesses match that of STEM-EDS measurements of the barrier thickness shown in the main text. As discussed in the main text, more realistic models of barriers are needed before accurate thickness values can be read out from these Monte-Carlo simulations.

\bibliography{bibliography}

\newpage

\appendix